\documentclass[fleqn,usenatbib]{mnras}

\usepackage{newtxtext,newtxmath}

\usepackage[T1]{fontenc}
\usepackage{ae,aecompl}

\usepackage{graphicx}	
\usepackage{amsmath}	
\usepackage{xcolor}

\usepackage[nolist]{acronym}

\newcommand{\nbody}{$N$-body}
\newcommand{\AMUSE}{{\tt AMUSE}}
\newcommand{\Ekster}{{\tt Ekster}}
\newcommand{\MSun}{\mbox{M}_{\odot}}

\newacro{sph}[SPH]{smoothed-particle hydrodynamics}
\newacro{amuse}[{\tt AMUSE}]{the Astrophysical Multipurpose Software Environment}
\newacro{gmc}[GMC]{Giant Molecular Cloud}

\let\oldpageref\pageref
\renewcommand{\pageref}{\oldpageref*}

\title[Early evolution of star clusters in spiral galaxies]{The formation and early evolution of embedded star clusters in spiral galaxies}

\author[S. Rieder et al.]{
Steven Rieder,$^{1,2}$\thanks{E-mail: steven@stevenrieder.nl (SR)}
Clare Dobbs,$^{1}$
Thomas Bending,$^{1}$
Kong You Liow,$^{1}$
\newauthor
and James Wurster$^{1,3}$
\\
$^{1}$ School of Physics and Astronomy, University of Exeter, Stocker Road, Exeter EX4 4QL, United Kingdom\\
$^{2}$ RIKEN Center for Computational Science, 7-1-26 Minatojima-minami-machi, Chuo-ku, Kobe, 650-0047, Hyogo, Japan\\
$^{3}$ Scottish Universities Physics Alliance (SUPA), School of Physics and Astronomy, University of St. Andrews,\\ North Haugh, St Andrews, Fife KY16 9SS, UK
}

\date{Accepted 2021 November 18. Received 2021 November 15; in original form 2021 September 22}

\pubyear{2021}

\begin{document}
\label{firstpage}
\pagerange{\pageref{firstpage}--\pageref{lastpage}}
\maketitle

\begin{abstract}
We present Ekster, a new method for simulating star clusters from birth in a live galaxy simulation that combines the \ac{sph} method Phantom with the $N$-body method PeTar.
With Ekster, it becomes possible to simulate individual stars in a simulation with only moderately high resolution for the gas, allowing us to study whole sections of a galaxy rather than be restricted to individual clouds.
We use this method to simulate star and star cluster formation in spiral arms, investigating massive GMCs and spiral arm regions with lower mass clouds, from two galaxy models with different spiral potentials.
After selecting these regions from pre-run galaxy simulations, we re-sample the particles to obtain a higher resolution.
We then re-simulate these regions for 3 Myr to study where and how star clusters form.
We analyse the early evolution of the embedded star clusters in these regions.
We find that the massive GMC regions, which are more common with stronger spiral arms, form more massive clusters than the sections of spiral arms containing lower mass clouds.
Clusters form both by accreting gas and by merging with other proto-clusters, the latter happening more frequently in the denser GMC regions.

\end{abstract}




\section{Introduction}
\label{introduction}
Stars form in galaxies, from collapsing molecular clouds \citep{ladalada}.
Generally, star formation doesn't form stars one at a time, but whole clusters and associations are born at similar epochs before feedback halts star formation \citep{pzmg10}.
We see this happen mostly on the scales of individual \ac{gmc}s or molecular cloud complexes (e.g. Carina \citep{preibisch11,buckner19}, Sco-Cen \citep{dezeeuw99,pm16,wm18}) and the arms of spiral galaxies \citep[e.g. in M51,][]{bastian05,scheepmaker09}.

Simulations of star clusters generally focus on either starting from a single spherical or elongated cloud, allowing for moderate to high resolution \citep[e.g.][]{bate2003,fpz15,Liow20}, or skip the star formation stage and take a distribution of stars as their starting point, using a spherical \citet{plummer} or \citet{king66} distribution \citep[e.g.][]{aarseth74,heggiehut}, a fractal type distribution \citep[e.g.][]{allison10,dpp20,yu11}, or by using observed regions to inform the starting point of a simulation \citep{sills18}.
Such simulations either ignore the galactic environment completely, or include it only in rudimentary form, e.g. as a galactic tidal field.
Since changing galactic dynamics can have a major impact on the formation of stars \citep[e.g. due to colliding flows,][]{inutsuka15,dobbs20}, this is not ideal.

As well as dramatic events such as mergers, bars \citep[e.g.][]{sheth05,hirota14,emsellem15,vera16,diazgarcia20,maeda20}, and the accretion of gas clouds \citep{alig18} can affect the local if not global star formation rate in galaxies.
However spiral arms are the most common environment for star formation in galaxies, with the gas dynamics of spiral arms determining at least where star formation occurs in galaxies, and also potentially influencing how much star formation occurs.

Ideally, to simulate the formation of star clusters self-consistently one would run a full galaxy simulation with enough resolution to form individual stars.
Computational limits make such a simulation unfeasible however.
Simulations that resolve the formation of individual stars \citep[][]{bate2003,bate12} are done on the scale of individual molecular clouds, but not necessarily even a whole cloud.
Simulations focusing on larger scales \citep[e.g.][]{bending20, sb17, tress21} therefore simplify the stellar population into `sink particles', whose properties are calculated using sub-grid physics.
These simulations ignore the dynamical evolution of the stellar population, and don't allow us to study the evolution of the star clusters. More recently though, approaches have been made to more fully represent stellar populations with individual star particles \citep[e.g.][]{hubber13,wall19,hirai21} even if the gas resolution is not correspondingly high, rather than simply using sinks whose masses constitute clusters or sub-clusters.

By following the full stellar population, this also allows us to study the dynamics of the stars. This has the advantage that it is possible to follow both the N-body dynamics plus the gas simultaneously, whereas typically previous simulations which follow cluster evolution using full N-body dynamics have assumed that the gas is expelled on some timescale and / or adopted a potential for the gas \citep{gb01,bg06,bk07,mb10,smith11,pfalzner13,bk13,farias15,shukirgaliyev18}.
Following the dynamics also allows mergers between clusters or sub-clusters to be fully resolved.
Previous spiral arm scale simulations found that cluster mergers are frequent \citep{sb17}, and simulations of both isolated \ac{gmc}s \citep{howard19,fpz15} and spiral arm \ac{gmc}s find mergers are important for massive cluster formation \citep{dobbs21}, but again these simulations typically used sinks representing clusters. Mergers have long been presumed to be important for the hierarchical growth of stellar clusters \citep{bonnell03,vazquezsemadeni17,grudic18,chen21}.

In this paper we follow cluster formation with full N-body dynamics in four different sections of spiral galaxies.
We start our simulations by extracting a section centred on a \ac{gmc} and a section of a spiral arm from two simulated spiral galaxies, which differ in spiral arm strength.
We use two different spiral models, with different strength spiral arms as a means of examining the role of spiral arms.
Previous work has found that spiral arms do not typically make a large difference to the global star formation rate in numerical simulations galaxies \citep{dobbs11,pettitt17,tress20,kimwt20}, rather the gas is simply gathered together in the spiral arms \citep{elmegreen86,vogel88}.
Observations of nearby arms in our Galaxy suggest they do not have a significant role \citep{eden13,eden15,urquhart21}, however there is some recent observational evidence that spiral arms have some impact on star formation rates over larger galaxy samples \citep{yu21}.
\citet{colombo14} also find more massive, and strongly star forming GMCs in the spiral arms of M51 compared to the inter-arm regions.
Furthermore whilst the previous numerical studies do not find large global changes, have found that particularly massive \ac{gmc}s can form in the presence of stronger, or tidally induced spiral arms \citep{dobbs11,pettitt18} compared to weaker or flocculent spiral arms.
Small scale models of colliding flows also find that strongly converging flows lead to massive clusters \citep[][]{dobbs20,Liow20}.
Such conditions would more likely occur in galaxies with stronger spiral arms, or locations in galaxies where localised strongly converging flows occur \citep{eden12,motte14}.
\citet{dobbs21} test the latter.
Here we test the former scenario where we simply vary the arm potential, which in turn produces stronger velocity gradients.

We aim to simulate the formation and early evolution of the full stellar population that would form in a section of a spiral galaxy by means of a multi-scale simulation.
To this end, we use \ac{amuse} \citep{10.1088/978-0-7503-1320-9, 2013A&A...557A..84P, 2009NewA...14..369P} to combine \ac{sph} with multi-scale \nbody{} dynamics and stellar evolution in a new simulation method, which we name \Ekster.
With this method we can simulate the formation of individual stars, while it also allows us to take the galactic environment into account.
This method is similar in approach to the methods used in the {\tt Torch} \citep[][]{wall19} and {\tt SIRIUS} \citep{hirai21,fujii21a} projects, though both focus on simulating individual clouds rather than a galactic environment - the former using grid-based hydrodynamics, the latter using an \ac{sph} method. 

The structure of this paper is as follows: in Section~\ref{method}, we introduce our simulation code, in Section~\ref{simulations} we discuss the initial conditions and settings used in our simulations, we describe the analysis of our simulations in Section~\ref{analysis}, in Section~\ref{results} we show the results from our simulations and discuss their implications and we conclude in Section~\ref{conclusions}.

\section{Simulation method}
\label{method}
To study how the galactic environment affects the formation and early evolution of star clusters, we want to simulate both a reasonably large section of a spiral galaxy and individual stars in the forming star clusters.
At the mass resolution required to form individual stars directly from gas \citep[e.g. $0.0011\MSun$ per SPH particle in][]{bate2003}, this is not feasible.
Our solution is to write a new simulation model that combines a hydro simulation at relatively low resolution (in this article: $1\MSun$ per \ac{sph} particle) with a method to form individual stars from star forming regions (``sinks'').
We implement this in the \Ekster{} simulation model, which we make publicly available via \url{https://github.com/rieder/ekster} \citep{ekster}.

\Ekster{} is a modular star formation simulation code that combines gas hydrodynamics, stellar dynamics and stellar evolution with a star formation method.
It also supports external tidal fields and feedback processes.
\Ekster{} employs \AMUSE{} as the environment that combines these elements.
In this article, \Ekster{} uses {\tt Phantom} \citep[][]{phantom} for gas hydrodynamics, while stellar dynamics is done with the high-precision, high-performance Tree/direct N-body hybrid code {\tt PeTar} \citep[][]{petarcode} and stellar evolution is done with {\tt SeBa} \citep[][]{seba1, seba2}.
We couple the gravitational dynamics between stars and gas  using {\tt Bridge} \citep{bridge}, an \AMUSE{} module based on \citet{2007PASJ...59.1095F}.
Time-dependent tidal fields are supported using the method described in \citet{rieder13}, and feedback from stellar winds and supernovae is an option via {\tt stellar\_winds.py} \citep[][]{stellarwindpy}.
Support for other types of feedback (e.g. radiation) is work in progress.

Our approach has several advantages over using a single hydrodynamics code (e.g. {\tt Phantom}) for this kind of simulation, primarily that we can support many more star particles.
{\tt Phantom} is designed to support sink particles in addition to gas, using direct N-body for the sinks and a $k$d-tree mechanism closely following \citet{gaftonrosswog11} for the gas to integrate gravity.
In our simulations, we expect the number of stars to be comparable to the number of gas particles, which would make this a very slow approach.
{\tt PeTar} instead uses a hybrid tree/direct N-body approach for integrating its particles \citep[based on the {\tt Pentacle} method from][]{iwasawa17}, combined with algorithmic regularisation for binary stars.
This approach is very useful for our simulations, as it will ensure that direct N-body gravity is used where needed in local groups (e.g. the cores of star clusters), while at larger distances tree gravity is still suitable.
Scaling is therefore substantially better than $N^2$.
Another advantage of our method is its modular nature, which means that we can easily add modules for e.g. feedback and switch out one code for another when desired.

The settings and parameters used for the different codes are listed in Table~\ref{tab:simulation_parameters} and described in the following sections.

\subsection{Gas}
Gas hydrodynamics in \Ekster{} is implemented via the \ac{sph} code {\tt Phantom} \citep{phantom}, though other hydrodynamical methods with an \AMUSE{} interface could be used in its place. 
Gas particles in {\tt Phantom} are integrated on individual time steps, while a global synchronisation timestep of ${\rm dt}_{\rm Phantom}$ is used (see Table~\ref{tab:simulation_parameters}).
By necessity, all gas particles in Phantom have an equal mass.
When a star formation region forms, gas particles accreted by this region are removed from {\tt Phantom}.
Conversely, gas can be added to {\tt Phantom} due to stellar feedback (stellar winds, supernovae).

\begin{table}
    \small
    \centering
    \caption{The parameters used for the simulations presented are listed above. dt$_{\rm bridge}$, dt$_{\rm Phantom}$ and dt$_{\rm PeTar}$  are the global timesteps for the {\tt bridge}, {\tt Phantom} and {\tt PeTar} codes respectively, $r_{\rm out}$ is the switchover radius between the tree and direct N-body methods for stellar gravity, and $\rho_{\rm crit}$ are $r_{\rm accr}$ are the accretion density and radius.} 
    \begin{tabular}{l|l}
        Parameter & setting \\
        \hline
        dt$_{\rm bridge}$ & 0.0025 Myr \\
        dt$_{\rm Phantom}$ & dt$_{\rm bridge}/2$ \\
        dt$_{\rm PeTar}$ & dt$_{\rm bridge}/256$ \\
        $r_{\rm out}$ & 0.001 parsec \\
        $\rho_{\rm crit}$ & $10^{-18}$ g cm$^{-3}$ \\
        $r_{\rm accr}$ & 0.25 parsec \\
        EOS & isothermal \\
        T & 30 K \\
    \end{tabular}
    \label{tab:simulation_parameters}
\end{table}

\subsection{Star formation}
The star formation routine in \Ekster{} is done in a way similar to the one described in \citet{wall19}.
When gas reaches a specified critical density ($\rho_{\rm crit}$, see Table~\ref{tab:simulation_parameters}) and passes additional checks (listed below), a sink particle will form and accrete all gas particles within a specified accretion radius $r_{\rm accr}$.
At critical density, this will lead to star forming regions of approximately $200\MSun$, which is enough to probe the IMF without a dearth of high-mass stars \citep{wall19}.
The position and velocity of the sink particle are taken as the centre-of-mass and centre-of-mass-velocity of the gas particles, while the velocity dispersion of the gas is also saved as a property of the sink.
This sink will then start forming stars by drawing a random mass from a \citet{kroupa} initial mass function, creating a star only if its mass is still higher than the mass of the star.
If this is the case, the star will be placed at a random position within the accretion radius with a velocity drawn from a Gaussian distribution centred on the velocity dispersion of the sink.
The star's mass is subtracted from the sink's mass.
This process continues until the sink no longer has enough mass to form the next star, at which point the accretion radius of the sink is shrunk to keep the original density of the star forming region constant.
The sink then acts as an unfinished proto-star, which can still accrete mass and potentially form new stars.

We also implement a secondary method for star formation, in which a group of sinks can act together as a mass reservoir for star formation.
This method would be suitable when star-forming regions consisting of a single sink won't have enough mass to prevent a dearth of high-mass stars, which would be the case when a higher \ac{sph} particle mass resolution is used than in this article.
This method is not further used here, but will be described in detail in a forthcoming article (Liow et al., submitted).

To decide if a star forming region should form, at each timestep in the simulation we check if any gas particle has reached a density higher than the critical density $\rho_{\rm crit}$.
If this is the case, we subject this gas particle to additional checks.
A star forming region will form if:
a) the smoothing length of the gas particle is smaller than half the accretion radius,
b) the thermal energy is less than half the potential energy,
c) the rotational energy and the thermal energy combined are less than the potential energy,
d) the total energy of the gas within the accretion radius is negative.
These checks are similar to those in \citet{bate95} and \citet[][paragraph 2.8.4]{phantom}.
The gas particles are processed in order of decreasing density, so any particle within the accretion radius of an already checked particle is not checked again.
Additionally, we add an option to skip these checks when any gas particle reaches a very high gas density of $N\times \rho_{\rm crit}$, in which case a star forming region will always be created to prevent the code from slowing down too much.
In this article, we set N to 10.

\subsection{Stars}
\label{sec:stellarevolution}
Once stars have formed, they are added to both the stellar evolution module (here: {\tt SeBa} \citep{seba1,seba2}) and the stellar gravity module (here: {\tt PeTar} \citep{petarcode}).
As the simulations in this article are run without stellar feedback and to a limited age, stellar evolution is used only to determine stellar properties, while the stellar masses are kept constant.
Stellar gravity is integrated using a combined tree/direct N-body method, with additional support for algorithmic regularisation to integrate binary stars.
We set the switchover radius from direct N-body to Tree gravity $r_{\rm out}$ to $0.001$ parsec.
Stars are integrated without any softening, allowing for the dynamical formation of binary stars.

\subsection{Coupling gravity}
We use the \AMUSE{} module Bridge \citep{bridge} to couple gravitational interaction between the gas particles and the stars.
We use a kick-drift-kick scheme for Bridge, in which particles are given a half-timestep kick, then drift for a full timestep, and are then given another half-timestep kick.
These kicks are calculated by the \citet{barneshut}-type tree code {\tt Fi} \citep{ficode}.
Generally, a tree code is sufficiently accurate in these calculations, since the internal gravity of the gas is also handled by a tree code.
The timestep for Bridge is set to 0.0025 Myr.

\subsection{Feedback}
{\tt Ekster} supports feedback in the form of stellar winds, supernovae and radiation.
Stellar winds and supernovae can be handled by the {\tt stellar\_winds.py} module \citep{stellarwindpy} in \AMUSE.
Similarly, for radiative feedback any of the \AMUSE{} modules for this can be used.
In this article however, we do not enable feedback as the mass resolution of the gas makes this impractical - a stellar wind particle can only be created once a star has released a mass equal to that of one gas particle as wind.
This would make it very impractical to use this method.
We will discuss further simulations with feedback enabled in a future article (Rieder~et~al.,~in~prep.).

\section{Simulations}
\label{simulations}
Our initial conditions are based on snapshots from each of two galaxy scale simulations, one of which is the simulation shown in \citet{dobbs13}.
The other is identical apart from the spiral potential used, and ran to provide initial conditions for this work.
The dimensions and locations of the regions we extract for our cluster simulations are shown in Table~\ref{tab:simulations}, as well as the mass resolution of our re-simulations.
From each simulation, we take two regions (see also Figures~\ref{fig:standard_model} and \ref{fig:strong_model}).
One region, which we denote `cloud', centres on a massive GMC, whilst the other region, which we denote `arm', centres on a section of spiral arm with a number of lower mass clouds.
Thus we have two `arm' regions, one in each simulation and two `cloud' regions, again one in each simulation.

\begin{table}
    \small
    \centering
    \caption{Details of the simulated regions. The `cloud' simulations start with a circular region focusing on a single massive cloud, while the `arm' simulations start with a square region which contain several smaller clouds. The `standard' runs use the galactic simulation from \citep{dobbs13} whilst the `strong' models use a galactic simulation with a stronger potential.}
    \begin{tabular}{l|l|l|l|l}
        Name & Particle mass & X & Y & Width \\
             & ($\MSun$) & (kpc) & (kpc) & (pc) \\
        \hline
        standard-cloud & 1.0 & -2.025 & 2.870 & 300\\
        strong-cloud & 1.0 & -5.225 & -1.050 & 300\\
        standard-arm & 1.0 & -1.800 & -1.800 & 600\\
        strong-arm & 1.0 & 2.500 & 0.500 & 600\\
    \end{tabular}
    \label{tab:simulations}
\end{table}

\subsection{Galaxy simulations}
\label{sec:galaxysims}
\begin{figure*}
    \centering
    \includegraphics[width=\textwidth]{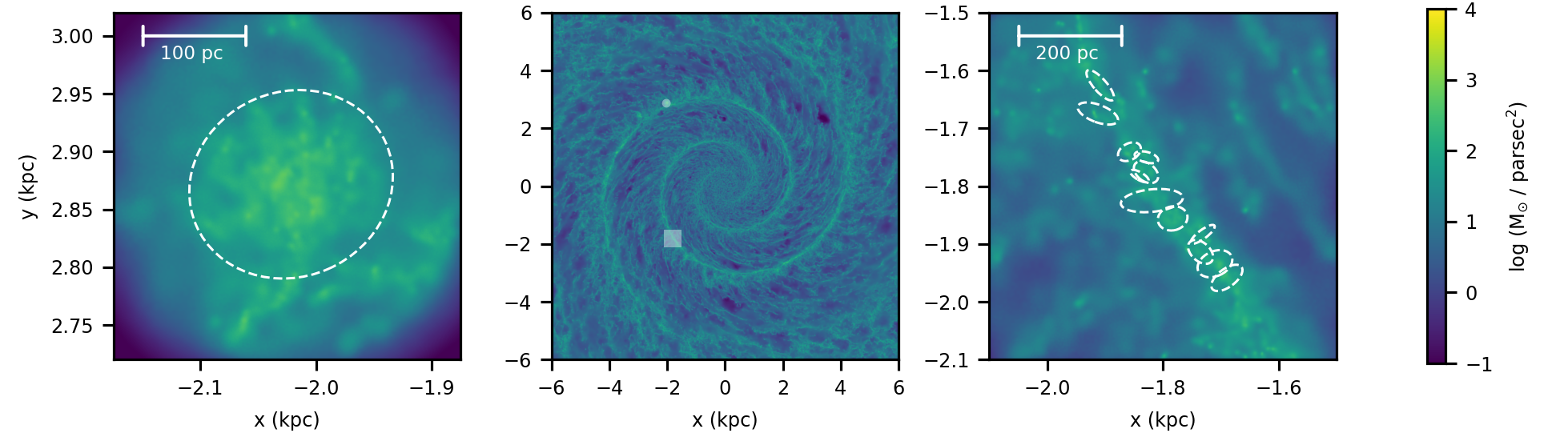}
    \includegraphics[width=\textwidth]{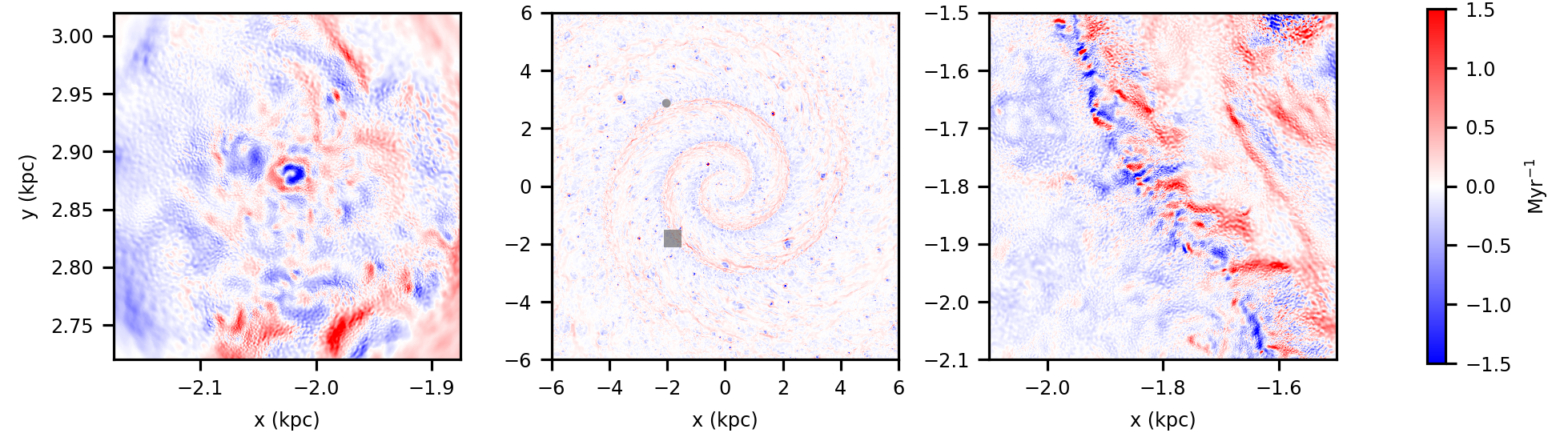}
    \caption{The column density (top panels) and velocity divergence (lower panels) are shown for the standard potential model. Left: standard-cloud area, centre: full galaxy with areas of interest highlighted, right: standard-arm area. The locations of the clouds are shown in the strong-cloud and strong-arm regions, for ease of plotting the shape of the clouds has been approximated to an ellipse along the major and minor axes of the cloud.}
    \label{fig:standard_model}
    \label{fig:standard_model_zoom}
\end{figure*}

\begin{figure*}
    \centering
    \includegraphics[width=\textwidth]{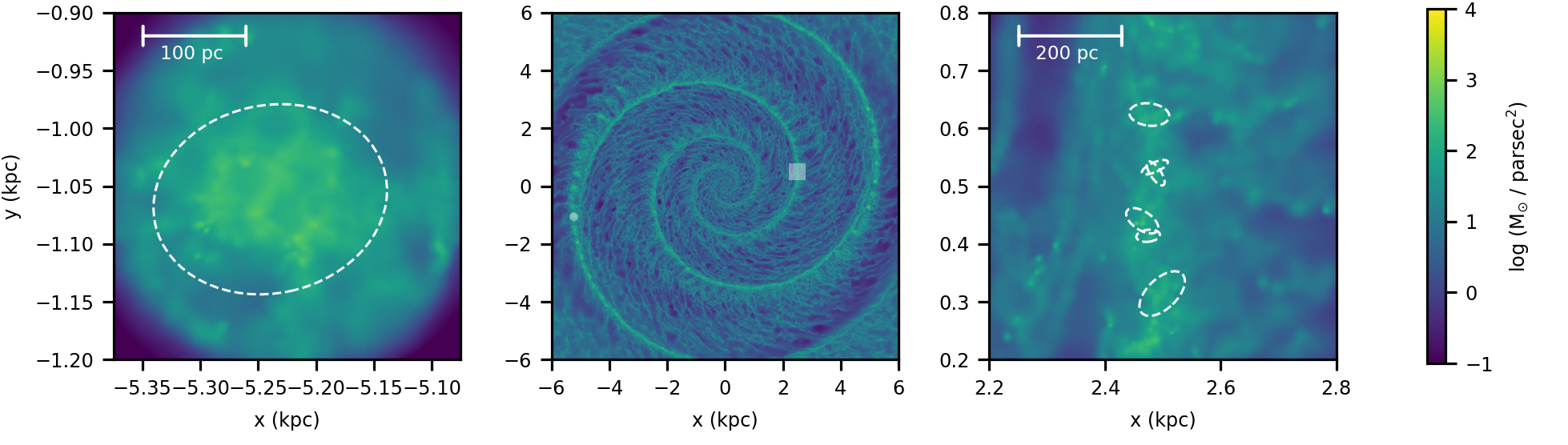}
    \includegraphics[width=\textwidth]{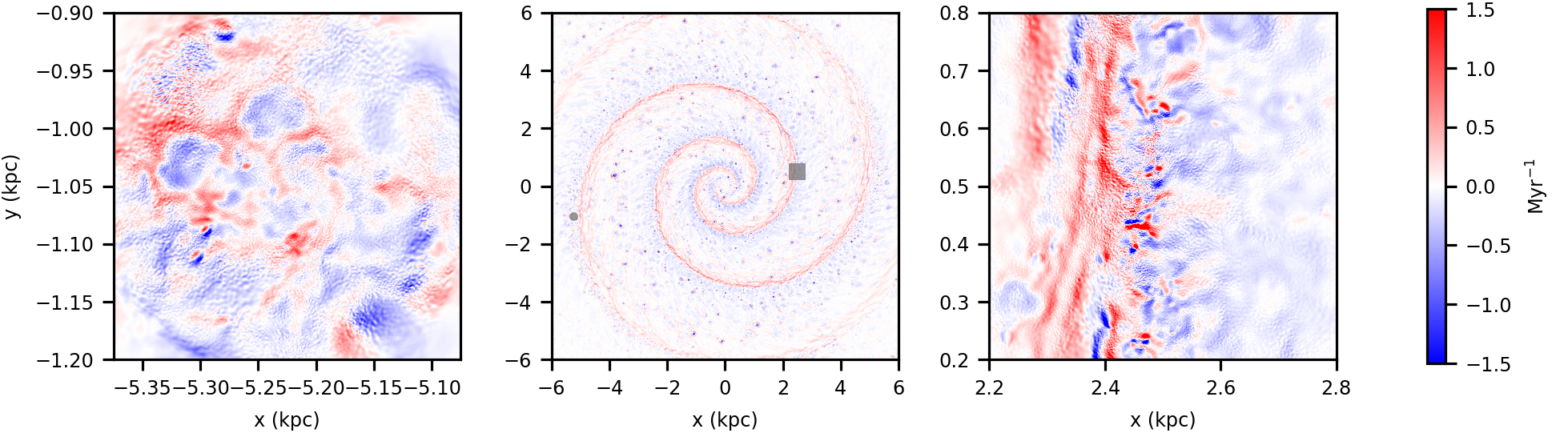}
    \caption{The column density (top panels) and velocity divergence (lower panels) are shown for the strong potential model. The model clearly shows stronger convergence in the spiral arms than the standard model. Left: strong-cloud area, centre: full galaxy with areas of interest highlighted, right: strong-arm area. The locations of the clouds are shown in the strong-cloud and strong-arm regions, for ease of plotting the shape of the clouds has been approximated to an ellipse along the major and minor axes of the cloud.}
    \label{fig:strong_model}
    \label{fig:strong_model_zoom}
\end{figure*}

The simulation taken from \citet{dobbs13} models the gas in a galaxy similar to the Milky Way, and adopts a logarithmic potential to produce a flat rotation curve \citep{galacticdynamics}, as well as a two-armed spiral perturbation following \citet{2002ApJS..142..261C}. 
The gas is initially assigned in a disc of radius 10 kpc with a uniform surface density of 8 M$_{\odot}$ pc$^{-2}$.
The simulation also includes heating and cooling \citep{glover07}, gas self gravity, and supernova feedback, which is applied instantly when stars are assumed to form.
Although relatively simple, the supernova feedback effectively disperses gas in molecular clouds, and leads to realistic cloud lifetimes, and cloud properties \citep{dobbs13}.
As the `standard model' for the re-simulations presented here, we take initial conditions from the simulation in \citet{dobbs13} (see Figure~\ref{fig:standard_model}).
However we also performed a second simulation (`strong model', see Figure~\ref{fig:strong_model}), which was identical to that shown in \citet{dobbs13}, except that we changed the form of the spiral potential.
We adapted the spiral potential so that the potential is twice as strong at half the radius of the galaxy, but drops away at the edge of the disc.
Explicitly, the spiral component from \citet{2002ApJS..142..261C} is multiplied by a factor
\begin{equation}
F(R)=\tanh(A*(R_a-R))+1    
\end{equation}
where $A=0.25$ and $R_a=6.2$ kpc are constants which determine the magnitude change in the spiral arm strength, and where the potential strength drops off with radius.
We use a $\tanh$ function as it provides a switch between $0$ and $1$.
For the `strong-cloud' model, the spiral arm potential is $\sim1.2$ times stronger, whilst for the `strong-arm' model it is $\sim1.7$ times stronger.
We used this potential so that we could increase the strength of the potential at radii of interest, where we select regions for our re-simulations, but not at the edge of the simulation where boundary effects tend to be more problematic.
The effect of using a stronger potential is to produce more bound, and more massive \ac{gmc}s (see also \citealt{dobbs11}). 
Numerous massive GMCs are visible just leaving the arms at radii of $\sim 5$ kpc in the galaxy simulation (Figure~\ref{fig:strong_model}, centre), whereas with the standard potential there are only one or two such GMCs.
These \ac{gmc}s also form earlier in the `strong' simulation.

We use the two different models to investigate how cluster formation depends on the different galaxy models, and the different morphologies of the \ac{gmc}s which form (see Figures~\ref{fig:standard_model} and \ref{fig:strong_model}).
A second reason for increasing the spiral potential is that a stronger potential should lead to higher converging flows.
We might predict that the clusters in the model with the stronger potential are more massive and / or form faster compared to the standard model.
In Figures~\ref{fig:standard_model} and \ref{fig:strong_model} we show the divergence of the velocity field for the two different models.
There is clearly stronger convergence in the spiral arms in the model with the stronger potential.
In terms of the velocities of the gas, the maximum velocity gradients are around 10--20 km s$^{-1}$ in the standard model across a 100 pc size region, compared to 20--30 km s$^{-1}$ with the stronger potential.

\subsection{Properties of the molecular clouds}

Before we show results from the simulations of cluster formation, we first determine the properties of the clouds which the clusters are born in. This is so that later we can relate the clusters which form to their natal clouds. 
We identify clouds in the original simulations using a friends of friends algorithm \citep{dobbs15}, which selects particles which are within a given distance of each other.
First, we select particles above a given density, which here is 1 cm$^{-3}$.
We then select particles which are within a distance of 5 pc of other particles, similar to previous work \citep{dobbs15,Liow20}. 

\begin{figure}
    \centering
    \includegraphics[width=\columnwidth]{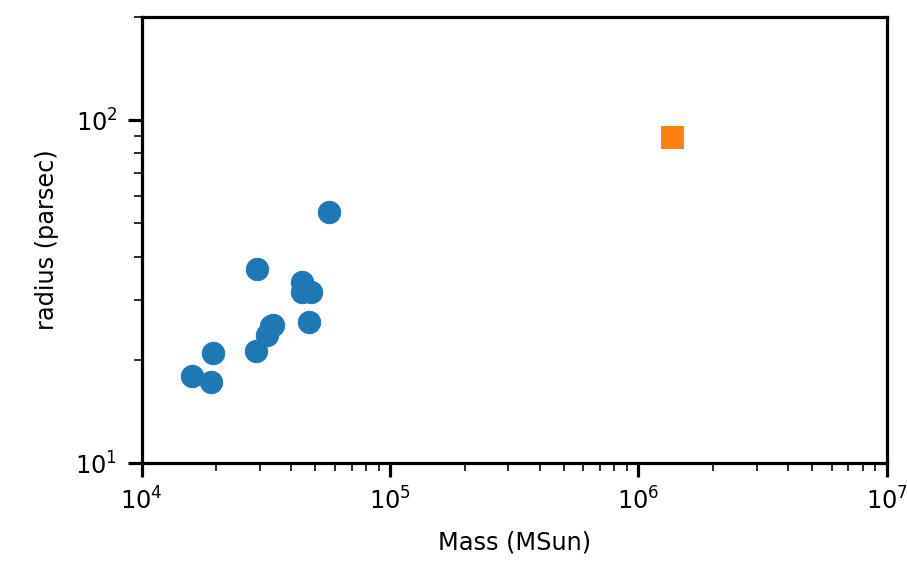}
    \includegraphics[width=\columnwidth]{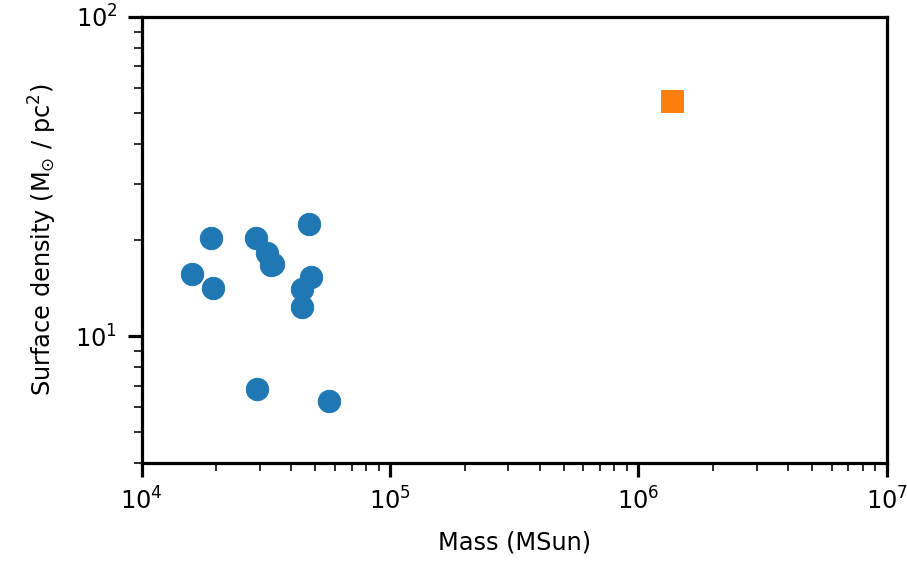}
    \includegraphics[width=\columnwidth]{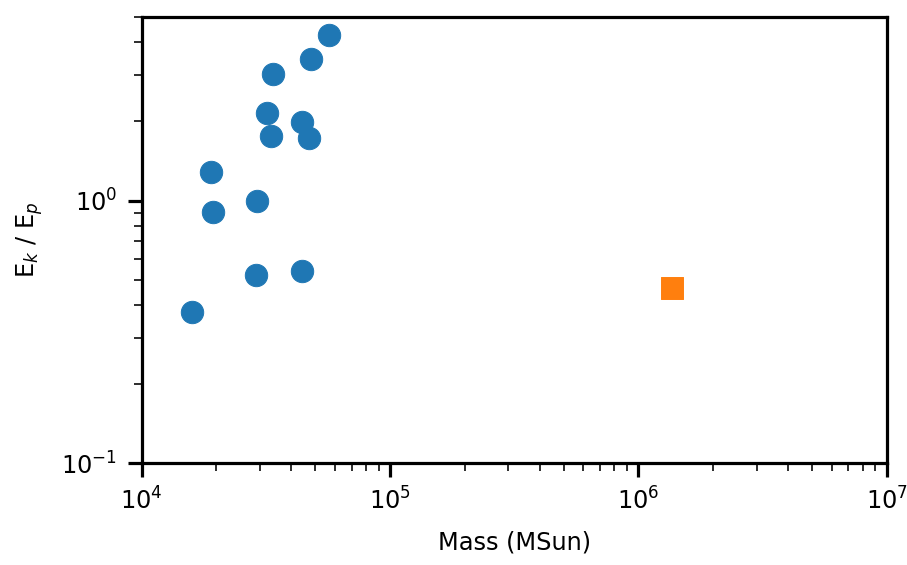}
    \includegraphics[width=\columnwidth]{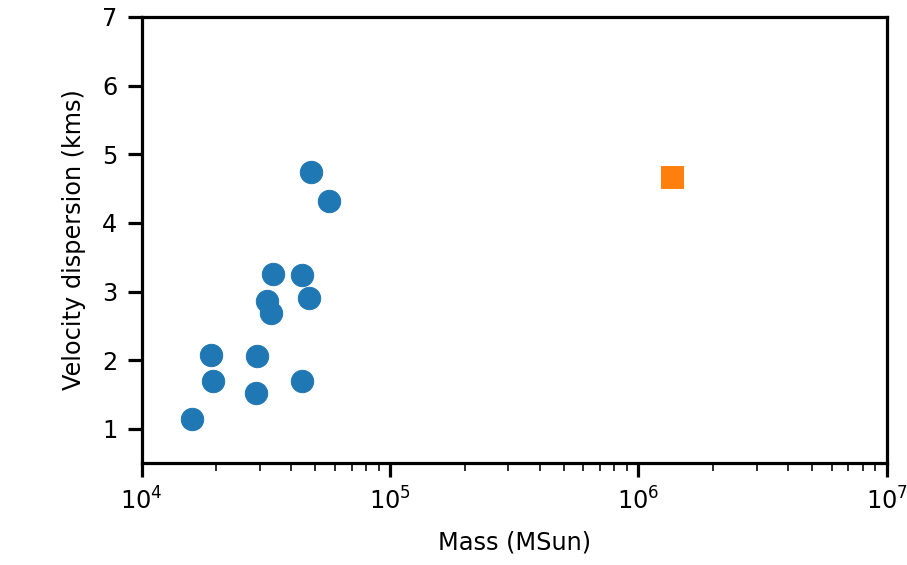}
    \caption{The radii, surface densities, kinetic divided by gravitational potential energy and velocity dispersion are plotted against mass for clouds in the `standard-arm' region (blue circles) and for the massive `standard-cloud' (orange square).}
    \label{fig:standardclouds}
\end{figure}

We show the properties of clouds along the spiral arm region (`standard-arm') and in the massive cloud (`standard-cloud') from the standard potential simulations in Figure~\ref{fig:standardclouds}.
We clearly see that the cloud from `standard-cloud' is an outlier compared to the `standard-arm' clouds, which is not surprising since the morphology of these areas is quite dissimilar.
The cloud from `standard-cloud' is an order of magnitude more massive, has a surface density which is around twice the clouds in the `standard-arm' region, and is strongly gravitationally bound.
The velocity dispersion is relatively high, this is likely from previous stellar feedback in or close to the clouds.

\begin{figure}
    \centering
    \includegraphics[width=\columnwidth]{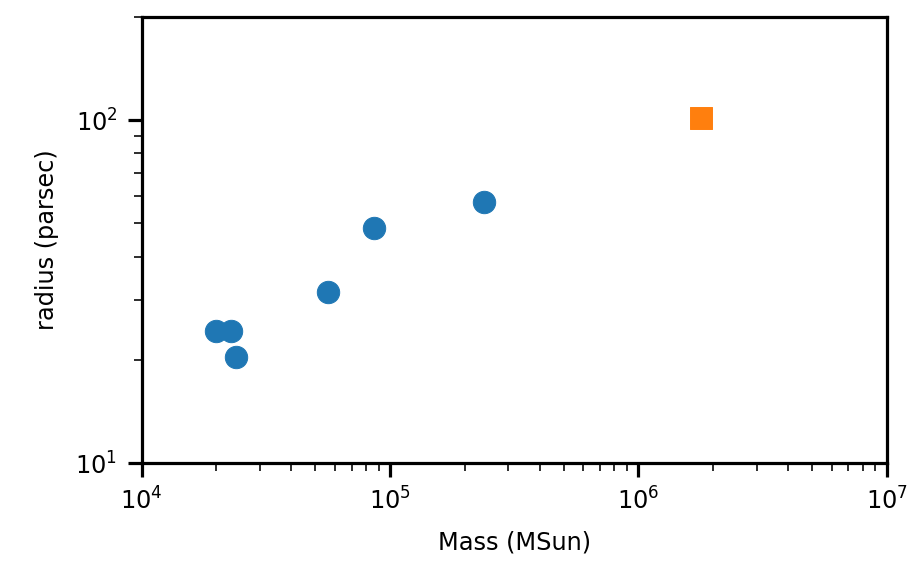}
    \includegraphics[width=\columnwidth]{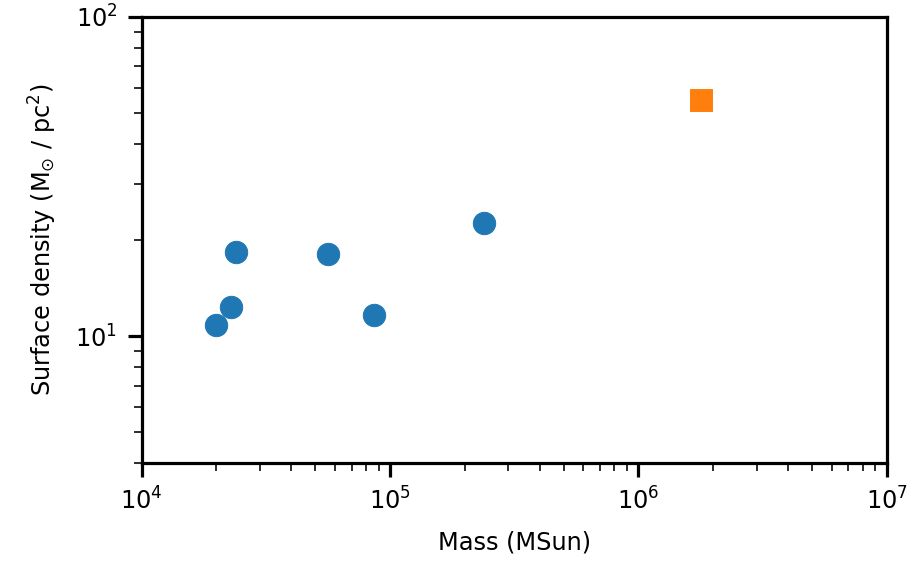}
    \includegraphics[width=\columnwidth]{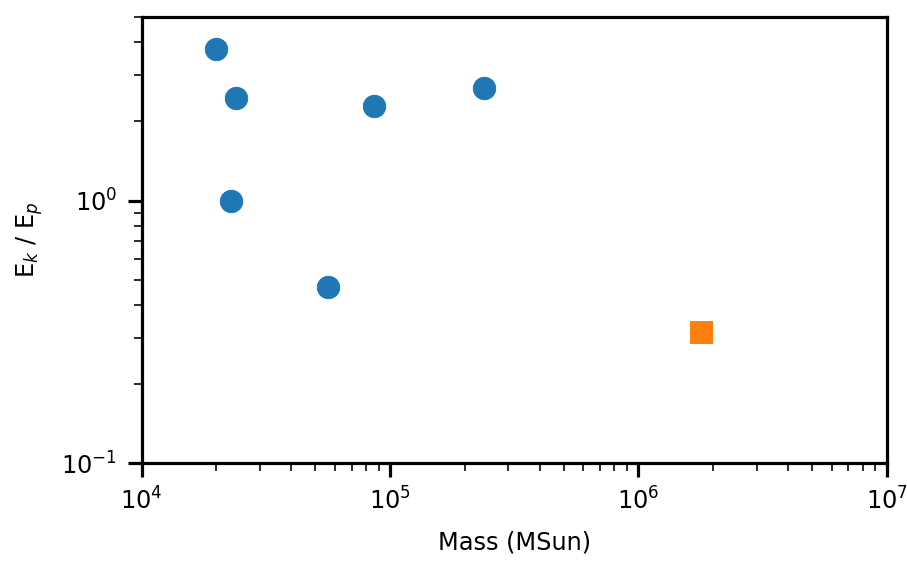}
    \includegraphics[width=\columnwidth]{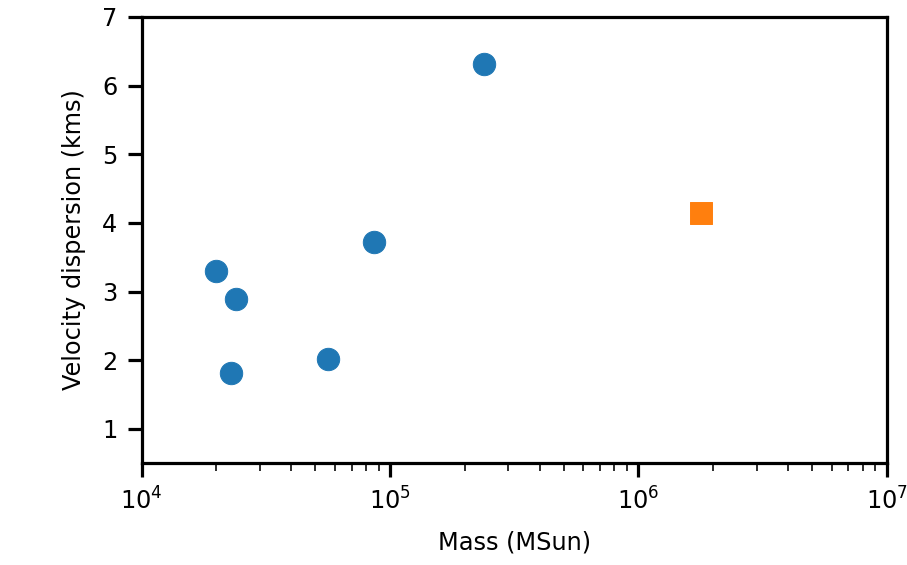}
    \caption{The radii, surface densities, kinetic divided by gravitational potential energy and velocity dispersion are plotted against mass for clouds in the `strong-arm' region (blue circles) and for the massive `strong-cloud' (orange square).}
    \label{fig:strongclouds}
\end{figure}

We show the properties of clouds for the strong potential model in Figure~\ref{fig:strongclouds}, again showing clouds from `strong-arm' and the massive cloud situated in `strong-cloud'.
Again the cloud in `strong-cloud' is an outlier, with a mass typically an order of magnitude higher than the clouds in the `strong-arm' region, and it has over twice as high a surface density. 
The cloud in the strong-cloud region is the most gravitationally bound, and is in fact more bound than the equivalent cloud in the `standard-cloud' region.
Again the cloud in `strong-cloud' has a relatively high velocity dispersion but it is not the highest compared with the clouds in the `strong-arm' region.

We also examined how typical these types of environment are in the global galaxy simulations.
For this we ran our friends of friends algorithm over the entire galaxy, and identify clouds of masses $>5\times10^5$ M$_{\odot}$, with a ratio of the kinetic and potential energy $E_k/|E_p|<2$, as similar to our ‘cloud’ models, and all other clouds similar to the clouds in our ‘arm’ models.
For the strong potential case, there are around 5 times as many clouds which satisfy this criterion compared to the standard potential. In terms of mass, these massive bound clouds constitute about 25\% of the mass in the strong potential galaxy, compared to 7\% of the mass for the standard potential galaxy.
This substantiates our claim that the stronger spiral potential and consequent stronger converging flows are resulting in more massive, bound clouds.

\subsection{Running the simulation}

After selecting the particles in our regions of interest (see Table~\ref{tab:simulations}), we re-sample the \ac{sph} particles following the method in \citet{bending20}.
Each original particle is split into 311 new particles of $1\MSun$ each.

We run simulations of each of these regions with {\tt Ekster}, using isothermal gas at 30~K.
To preserve the large-scale environment of the original galactic simulation, we include the same tidal field used in the galaxy simulations (see Section~\ref{sec:galaxysims}).

Since our simulations do not include feedback, we limit our simulations to the embedded phase of star cluster formation, i.e. up to 3 Myr \citep{ladalada,kimj21}.
We save a snapshot every 0.01 Myr.

\section{Analysis}
\label{analysis}
Once stars form, we use {\tt Hop} \citep{hop} to calculate the stellar density in every snapshot and to locate density peaks.
To ensure clusters are found in the same way in each simulation and snapshot, we explicitly set {\tt Hop} to use a relatively low threshold of $>3 \MSun / {\rm pc}^3$, an outer density threshold of $1 \MSun / {\rm pc}^3$ and a relative saddle density threshold factor of 0.5, while the number of neighbours to detect the local density is set to 64.
Using the density peaks, or cores, as a starting point, we then use a different algorithm to find clusters.
Starting from the densest peak, we calculate the radius for a sphere centred on the peak at which the average stellar density in the whole sphere becomes less than $10 \MSun / {\rm pc}^3$.
We consider all the stars in such a sphere to be part of the same cluster, and any further density peak residing within the cluster's radius will be considered a sub-cluster.
For each cluster, we determine its mass, velocity dispersion, half-mass and other Lagrangian radii, and the radius, velocity and density of the core.

We determine the formation history of a cluster by calculating the expected location of the cluster core in an earlier snapshot as $r_{s-1} = r_{s} - dt*v_{s}$, where $dt$ is the time between snapshots and $r_{s}$ and $v_{s}$ are the position and the velocity of the core at snapshot $s$.
We then search for the cluster nearest to this expected location that contains at least half of the cluster members that would have formed at this time.
This cluster is then designated the predecessor.

\section{Results and discussion}
\label{results}
In this section we present results from our simulations of cluster formation and evolution.
We compare the evolution of the four largest star clusters between similar regions in the two galaxies, as well as between different regions in the same galaxy.

\begin{figure}
    \centering
    \includegraphics{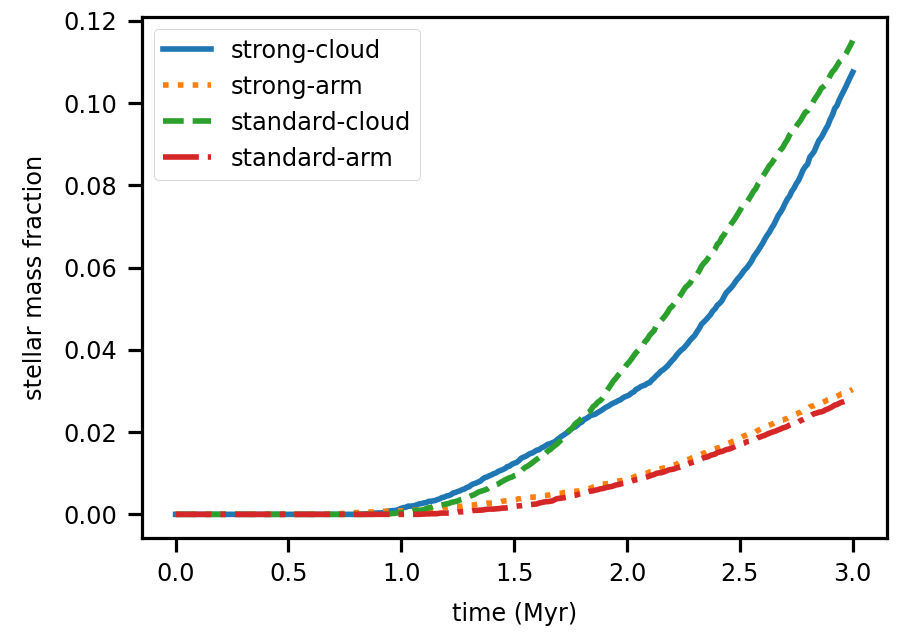}
    \caption{Stellar mass fraction over time in the four simulations. In the `cloud' simulations a larger fraction of the initial mass has been converted to stars than in the `arm' simulations.
    }
    \label{fig:sfr}
\end{figure}

\begin{figure}
    \centering
    \includegraphics{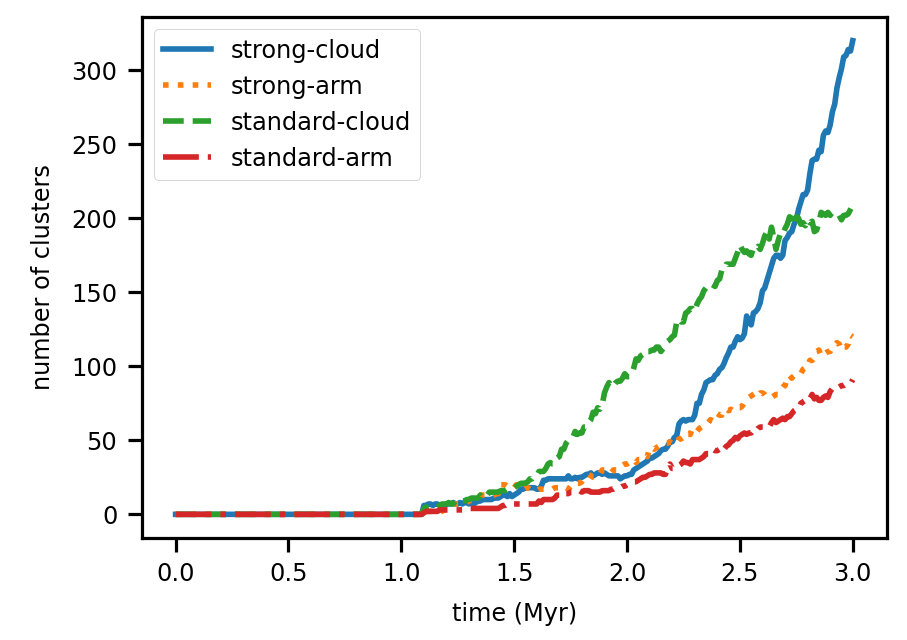}
    \caption{The number of clusters formed in the different models is shown over time.}
    \label{fig:n_clusters}
\end{figure}

We find that in all four simulations, star formation starts after $\sim$ 1 Myr (see Figure~\ref{fig:sfr}).
Both of the `strong' models produce a larger number of clusters than their `standard' equivalent by 3 Myr, although the `strong-cloud' model initially forms relatively few clusters (see Figure~\ref{fig:n_clusters}).
Star formation is strongest in the `cloud' regions of each simulation, but there is not a great difference between either the `cloud', or the `arm', regions from the `strong' and `standard' simulations. For the `cloud' simulations, over 10\% of the gas is converted to stars, whereas in the `arm' regions it is around 3\%, although we would expect that feedback may decrease these numbers.
The typical star formation efficiency of GMCs is 1--5\% (e.g. \citealt{ladalada}) but may be higher for more massive clouds or clusters \citep{vutis16,oschendorf17,tsuge21}. 

In Figure~\ref{fig:regions3myr} we show images of the column density of the four models in the final snapshot, with an inset showing the most massive cluster in each simulation. 
As expected, in the `cloud' models stars have formed primarily towards the centre of the clouds, in what appears by eye to be more massive clusters.  
This is not that surprising since the massive clouds selected in the `standard-cloud' and `strong-cloud' are strongly gravitationally bound, and we would expect gas to condense towards the centre (see Figures~\ref{fig:standardclouds} and \ref{fig:strongclouds}).
In the `arm' models, clusters are more spread out along the total length of the arm. Again, this is not surprising as for the `arm' models, several clouds are distributed along the spiral arm.
By eye, there is little obvious difference between the `strong' and `standard' models.

\subsection{Cluster evolution over time}
\begin{figure}
    \centering
    \includegraphics[width=0.93\columnwidth]{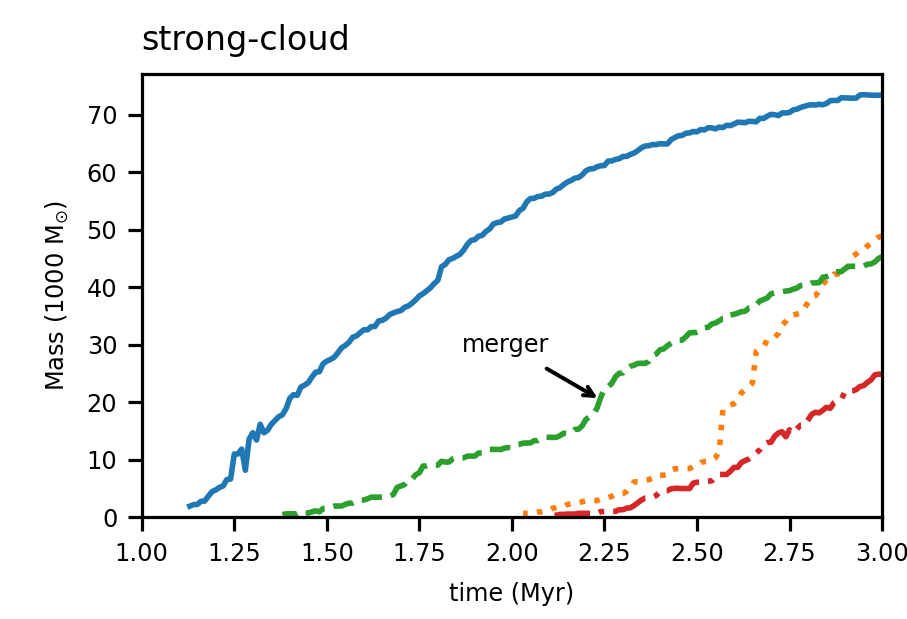}
    \includegraphics[width=0.93\columnwidth]{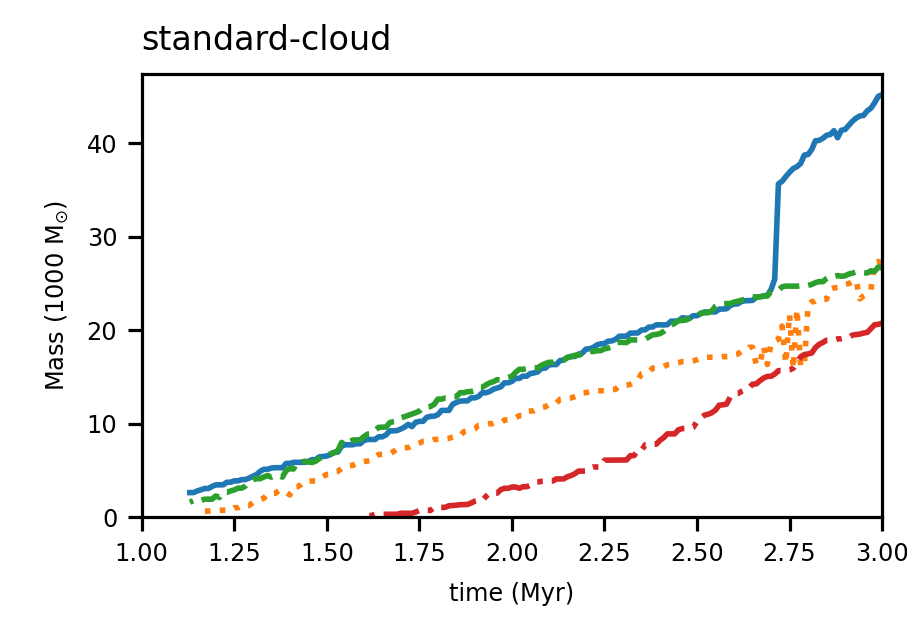}
    \includegraphics[width=0.93\columnwidth]{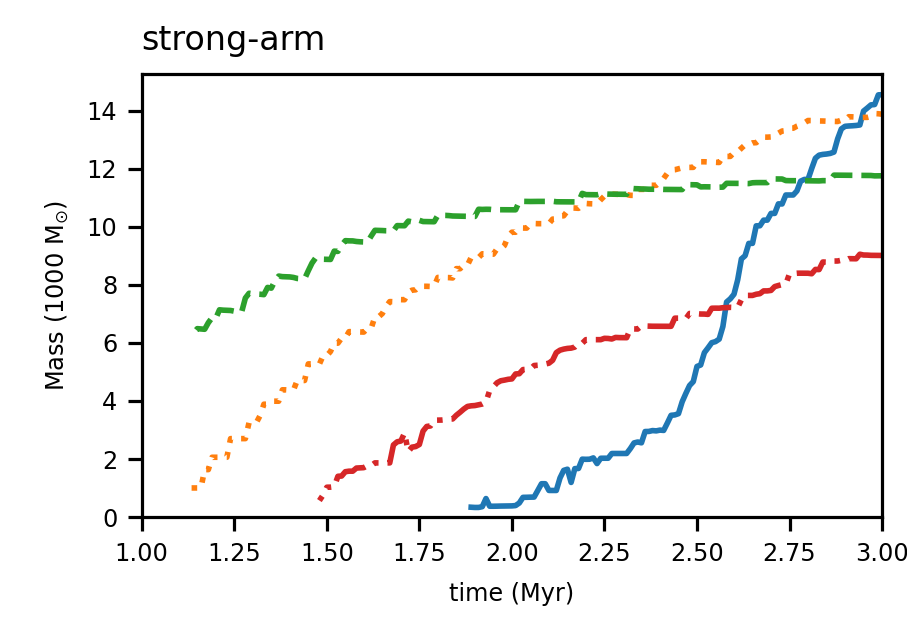}
    \includegraphics[width=0.93\columnwidth]{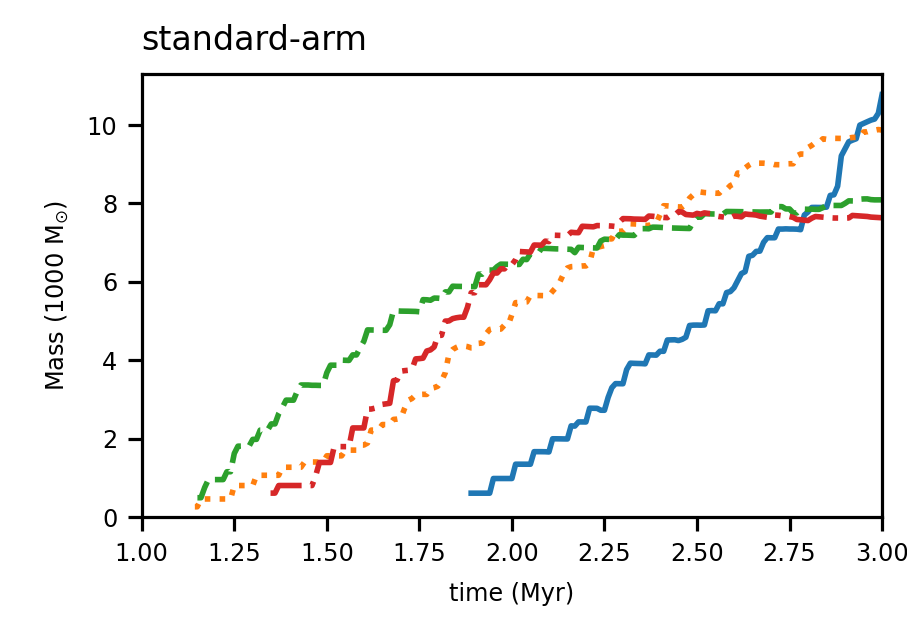}
    \caption{The mass of the four largest clusters is shown from the four different simulations.
    Clusters reach much higher masses in the `cloud' simulations, and in the strong-cloud simulation in particular.}
    \label{fig:massgrowth}
\end{figure}

\begin{figure}
    \centering
    \includegraphics[width=0.93\columnwidth]{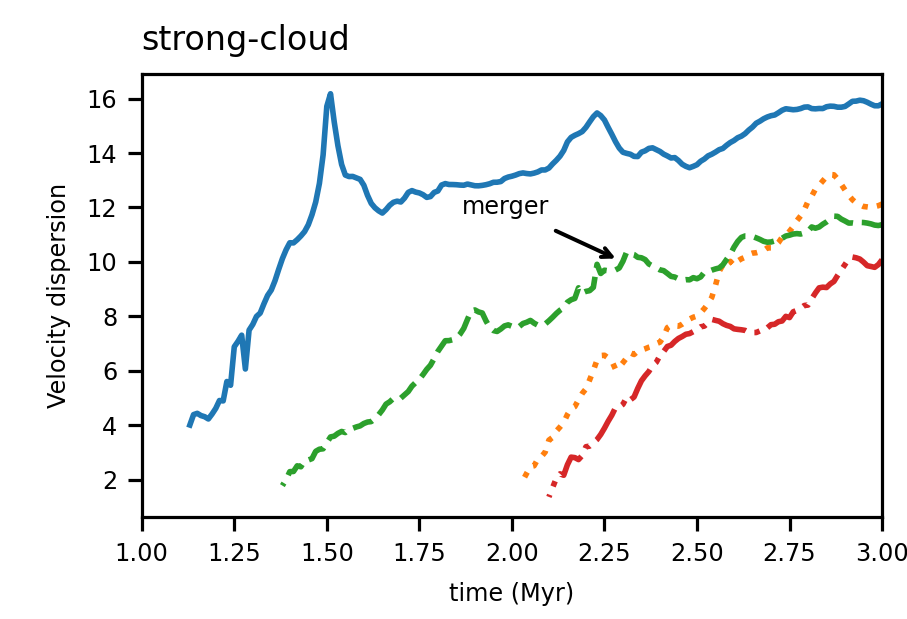}
    \includegraphics[width=0.93\columnwidth]{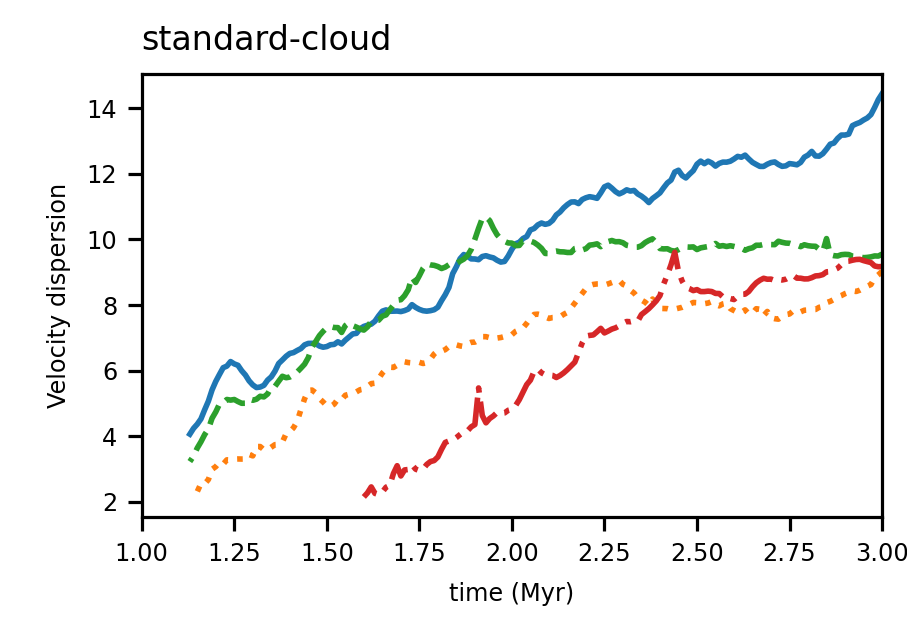}
    \includegraphics[width=0.93\columnwidth]{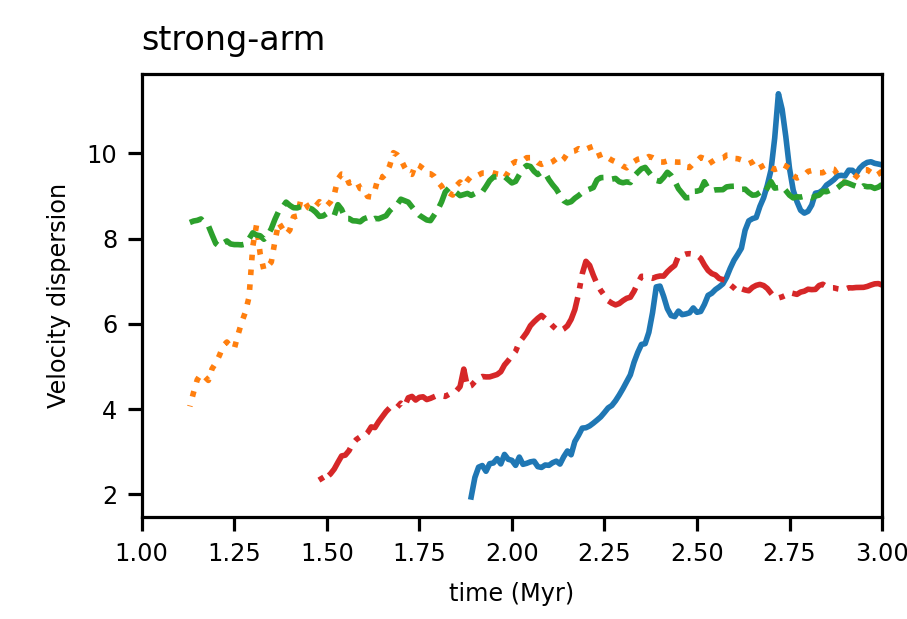}
    \includegraphics[width=0.93\columnwidth]{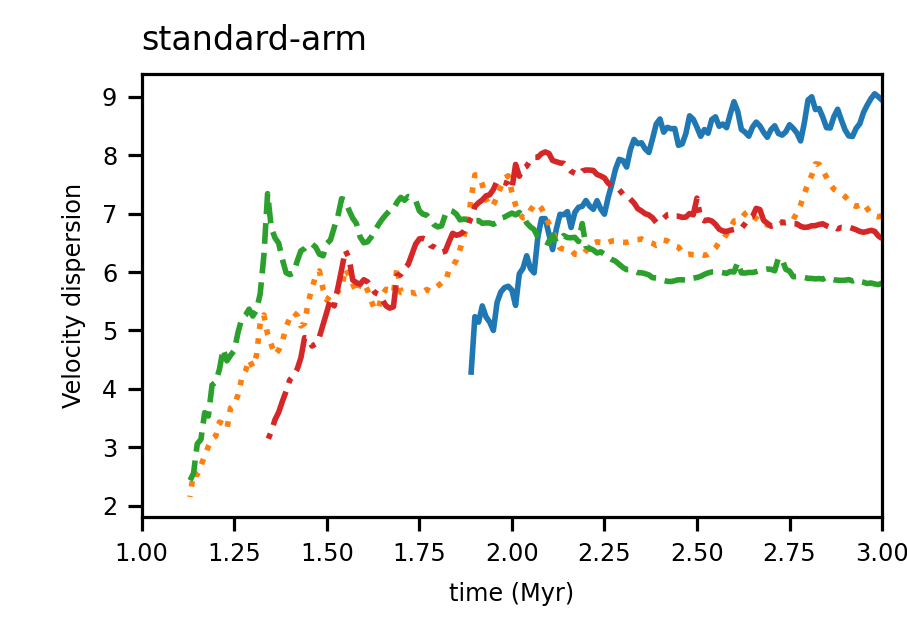}
    \caption{The velocity dispersion of the four largest clusters are shown for the four simulations versus time.}
    \label{fig:velocity_dispersion}
\end{figure}

\begin{figure}
    \centering
    \includegraphics[width=0.93\columnwidth]{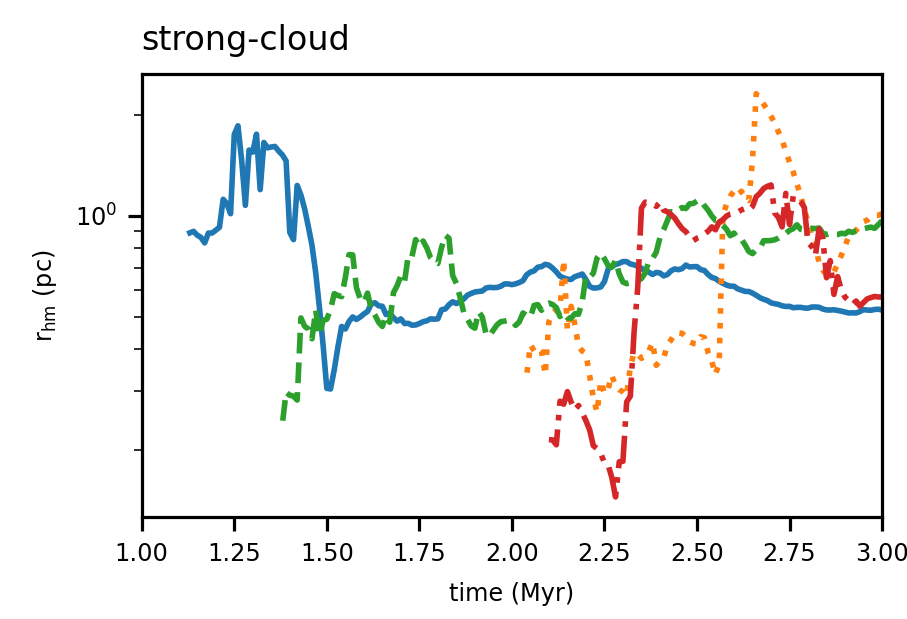}
    \includegraphics[width=0.93\columnwidth]{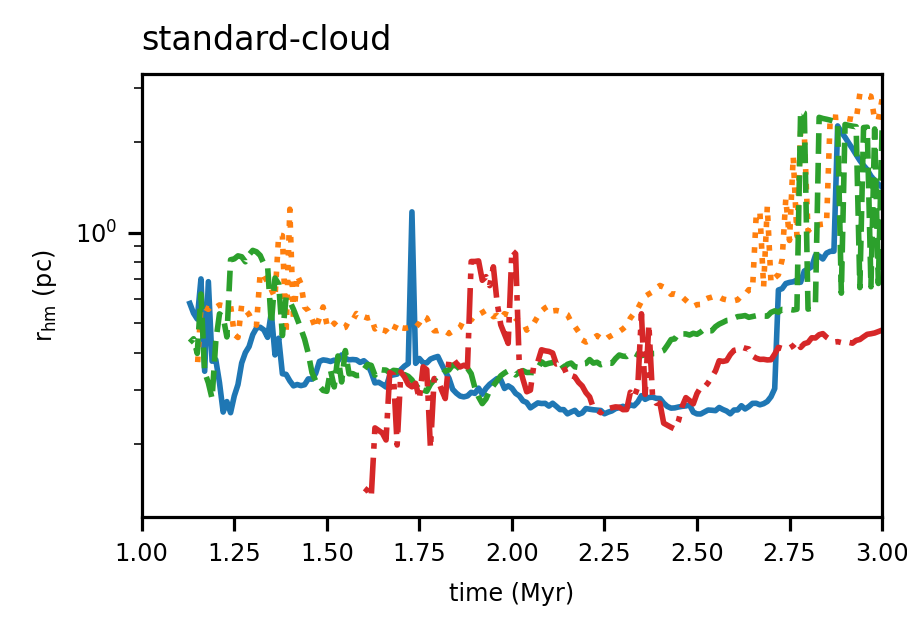}
    \includegraphics[width=0.93\columnwidth]{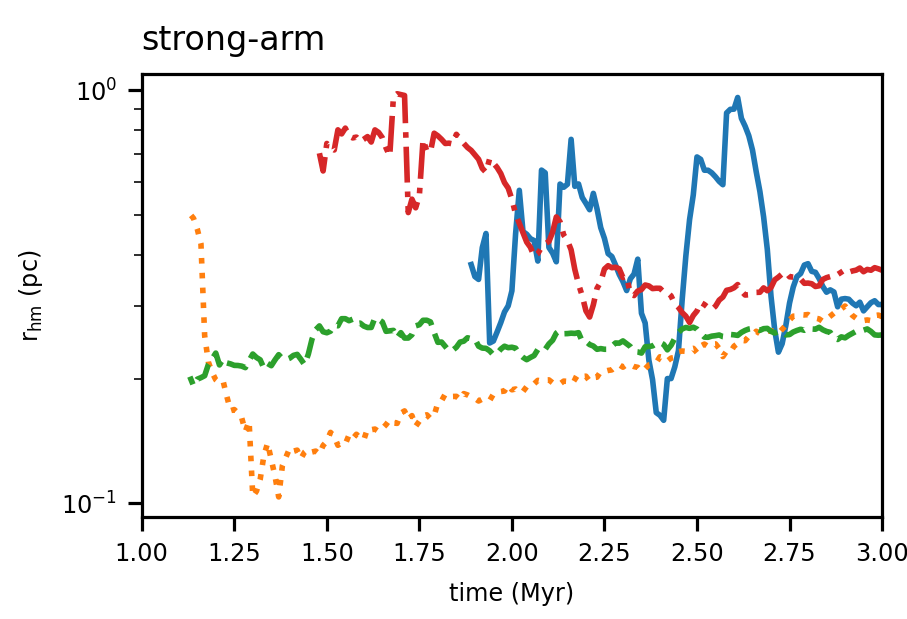}
    \includegraphics[width=0.93\columnwidth]{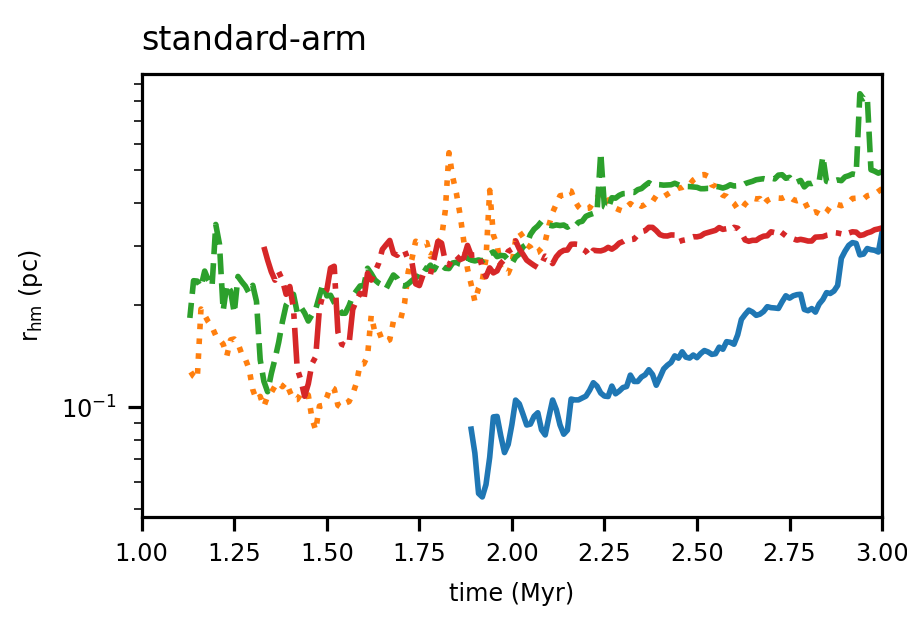}
    \caption{The half-mass radius of the four largest clusters for the four simulations are shown versus time.}
    \label{fig:rhm}
\end{figure}

We compare the evolution of the four largest star clusters between similar regions in the two galaxies, as well as between different regions in the same galaxy.
We show the mass (Fig.~\ref{fig:massgrowth}), velocity dispersion (Fig.~\ref{fig:velocity_dispersion}) and half-mass radius (Fig.~\ref{fig:rhm}) as they evolve over time.

Figure~\ref{fig:massgrowth} shows that cluster masses in the `cloud' models grow to considerably larger values than in the `arm' models.  
As we saw from Figures~\ref{fig:standardclouds} and \ref{fig:strongclouds} the initial mass reservoir of the clouds from which the clusters form in the `cloud' models is quite different from the `arm' models.
In the `cloud' models, there is $\sim10^6$ M$_{\odot}$ of strongly gravitationally bound gas, whereas in the `arm models', the clusters form in $\sim10^{4-5}$ M$_{\odot}$ clouds which may not even be bound. 

\begin{figure*}
    \includegraphics[width=\columnwidth]{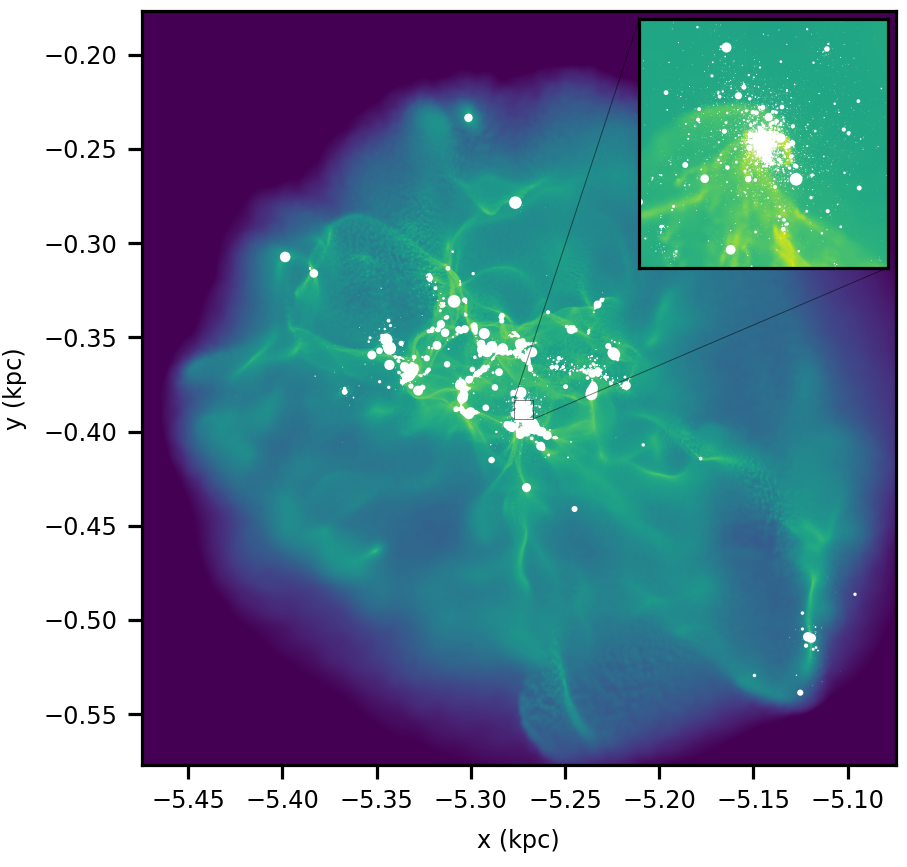}
    \includegraphics[width=\columnwidth]{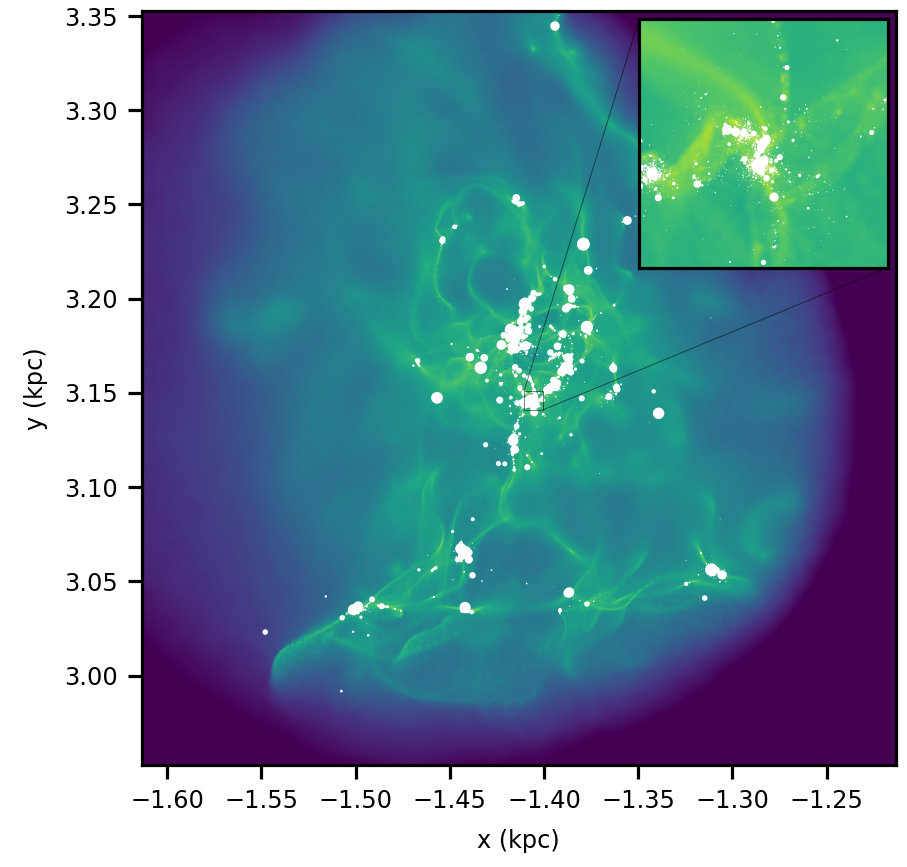}
    \includegraphics[width=\columnwidth]{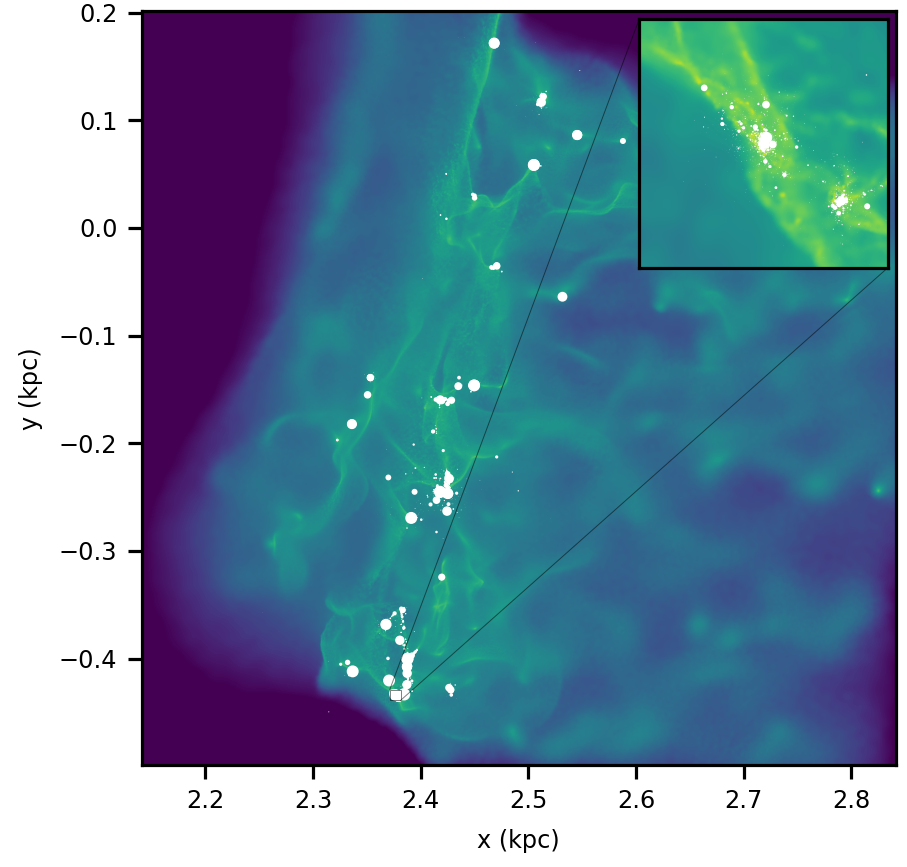}
    \includegraphics[width=\columnwidth]{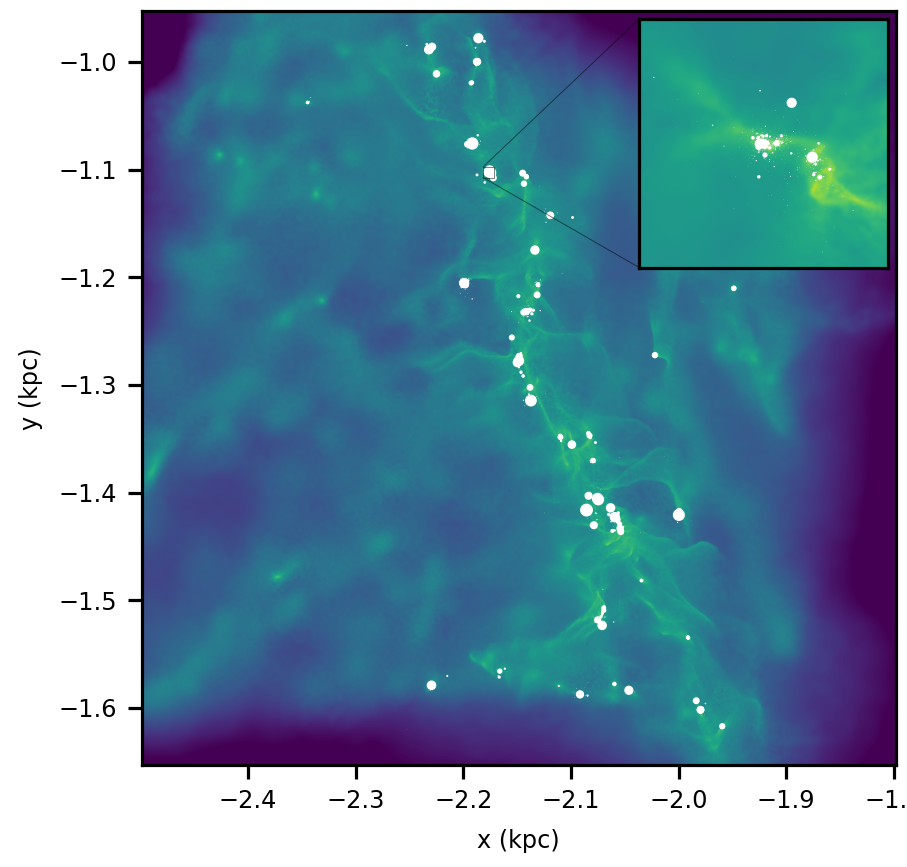}
    \caption{The four regions are shown at 3 Myr. 
    The inset regions are 10 parsec wide and zoom in on the largest star cluster of each simulation.
    The size of the stars is proportional to their radius. The column density scale is the same as Figures~\ref{fig:standard_model} and \ref{fig:strong_model}.
    Top row: strong-cloud (left) and standard-cloud (right).
    Bottom row: strong-arm (left) and standard-arm (right).}
    \label{fig:regions3myr}
\end{figure*}

Initially, the mass of the clusters increases linearly, indicating a steady accretion of gas by these regions which is converted into stars.
In some instances, the increase in mass is smooth, but in other cases (e.g. blue, green and orange lines, top panel), we see sudden jumps in the mass\footnote{The noise in the orange line, second panels, is due to our automated algorithm swapping between the progenitor cluster and a merging cluster, in most other cases the algorithm just selects the progenitor.}.
These happen when the core of one cluster enters the radius of another, leading to a merger of the two.
Such mergers of clusters happen more frequently in the `cloud' simulations than the `arm' simulations.
As we can see from Figure~\ref{fig:regions3myr}, top panels, insets, 
there are multiple smaller clusters close to the most massive clusters in the `cloud' simulations, and some of these will merge.  
Mergers are visible as a peak in the velocity dispersion, Fig.~\ref{fig:velocity_dispersion}, often correspond to a jump in mass, Fig.~\ref{fig:massgrowth}, and also as an increase in the half-mass radius, Fig.~\ref{fig:rhm}. We can see examples of features corresponding to a merger in the top panel of the mass and velocity dispersion plots (see e.g. blue lines in top panels, and explicitly labelled in the green lines). Typically we see an increase in mass which precedes the increase in velocity dispersion, as the two clusters come together spatially, and then the N-body dynamics drive an increase in the velocity dispersion.

\begin{figure*}
    \centering
    \includegraphics[width=\linewidth]{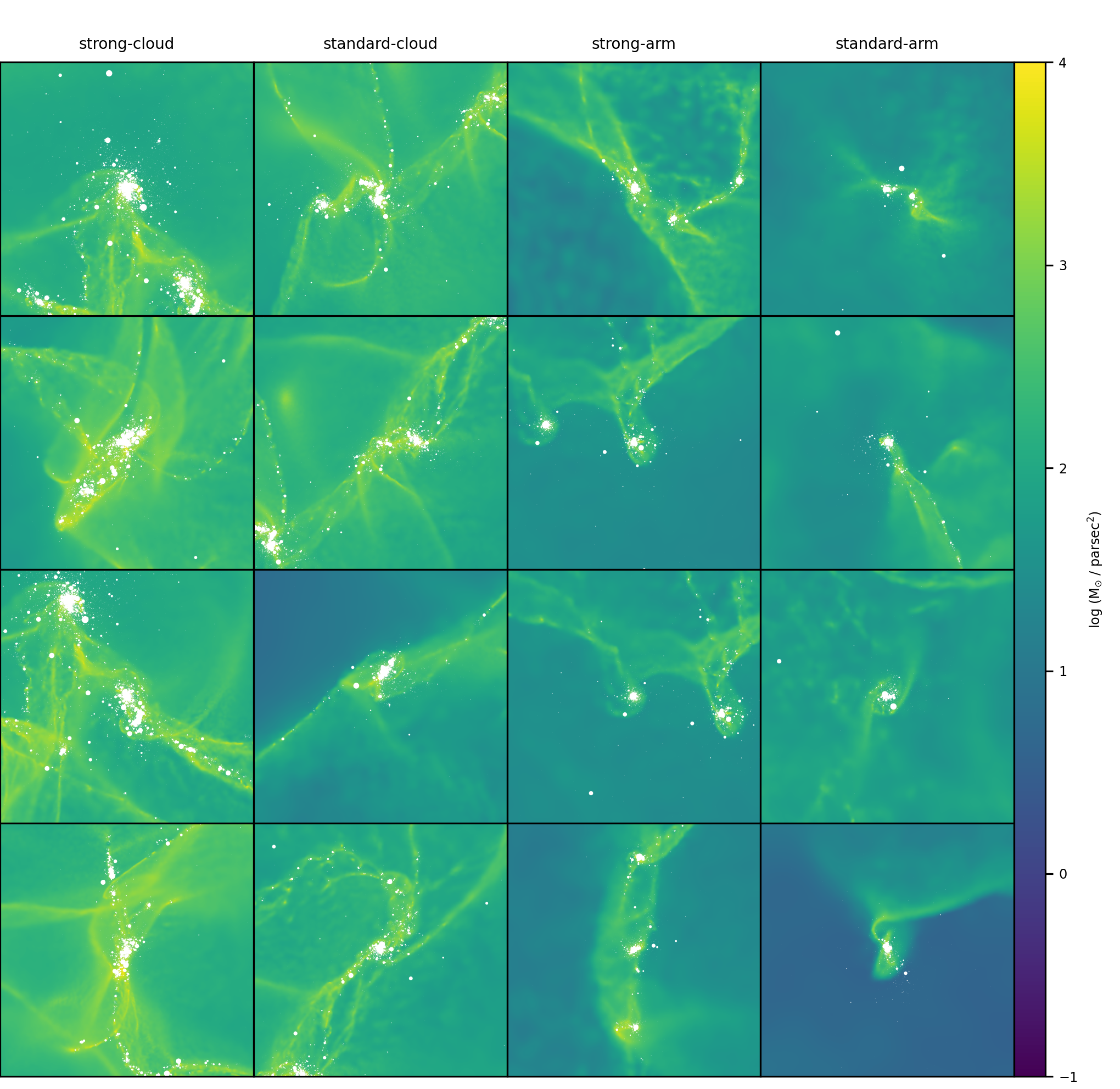}
    \caption{The four most massive clusters (decreasing in mass from top to bottom) from each simulation are shown at 3 Myr.}
    \label{fig:clusters_t0300}
\end{figure*}

In the `arm' simulations, growth slows down for most clusters after $\sim 1 {\rm Myr}$, while in the `cloud' simulations clusters continue to grow.
This is caused by the star clusters no longer being near a reservoir of gas, either because they have used up all the gas in their surroundings or because the stars have decoupled from the gas. 
We see the decoupling occur when the gas undergoes a shock from larger scale converging flows, whilst the cluster becomes displaced from the shock and is dominated by the smaller scale dynamics of the constituent stars (see also \citealt{renaud15}).
In the `cloud' simulations, this decoupling seems to not take place and as a result the clusters can grow larger.
Figure~\ref{fig:clusters_t0300} shows panels centred on each of the four most massive clusters, in each simulation, at a time of 3 Myr.
These panels show that for the `cloud' models, there is usually still high column density gas within at least a 10 pc radius of the cluster.
By contrast, for the `arm' models, the gas column density is lower, and in particular for the lowest mass `standard-arm' cluster, the vicinity of the cluster contains predominantly lower density gas. 
Together with the increased number of mergers, this results in the `cloud' simulations producing much more massive clusters than the `arm' simulations, in particular the `strong-cloud' simulation.
Again, from Fig.~\ref{fig:clusters_t0300} we also see that for the `cloud' models there are numerous, sometimes quite massive, clusters within 10 pc of the central cluster, indicative of the likelihood of mergers. 

\subsection{Cluster properties}
\begin{figure}
    \centering
    \includegraphics[width=0.9\columnwidth]{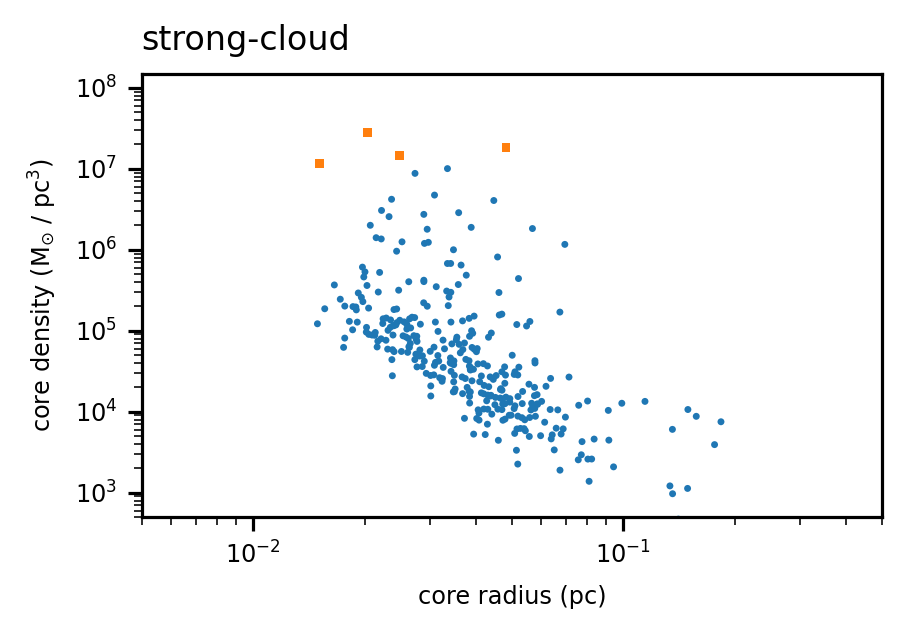}
    \includegraphics[width=0.9\columnwidth]{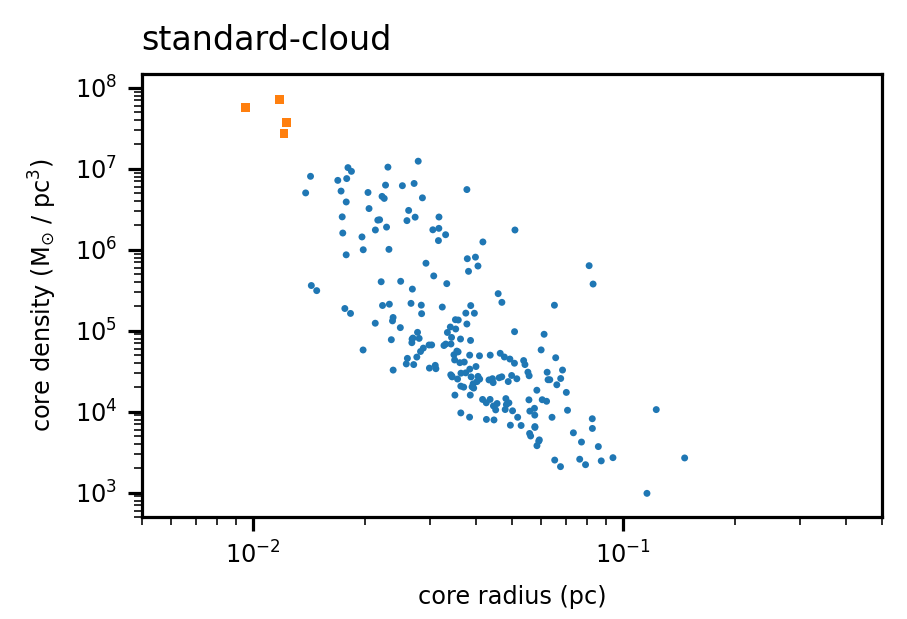}
    \includegraphics[width=0.9\columnwidth]{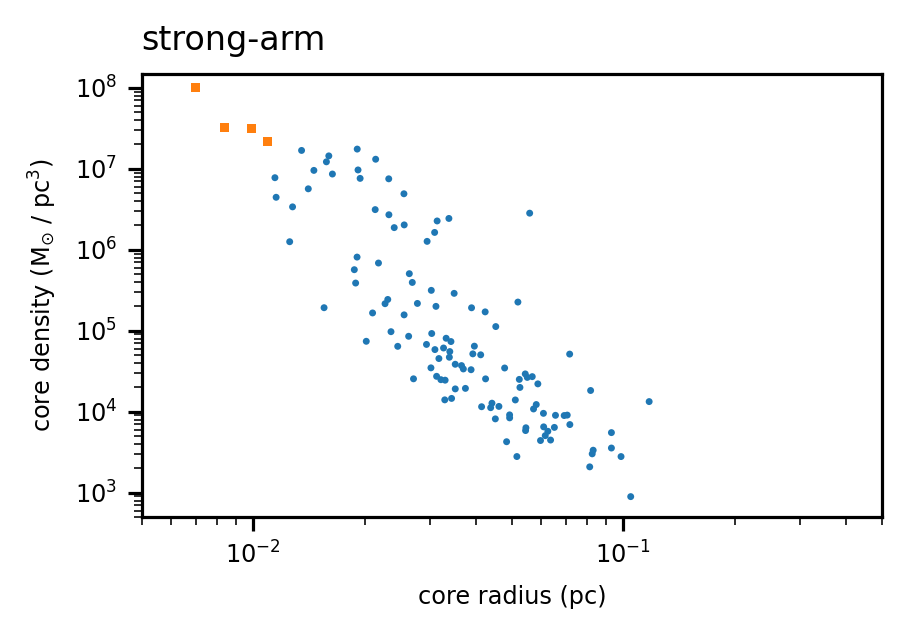}
    \includegraphics[width=0.9\columnwidth]{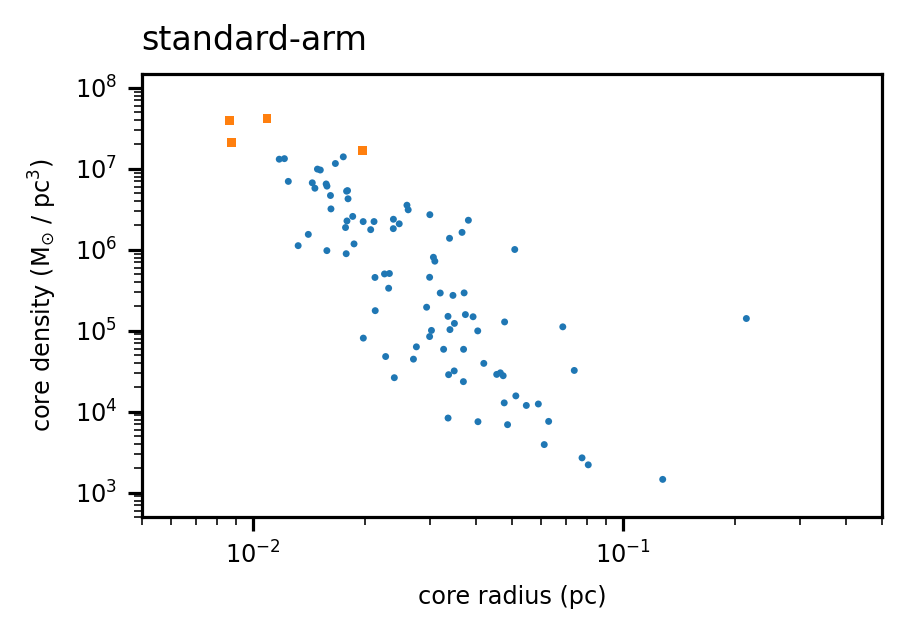}
    \caption{The core radius versus core density is shown for all clusters at 3.0 Myr.
    The clusters all show similar populations, but the `cloud' simulations yield more clusters and more massive clusters than the `arm' simulations.
    The most massive clusters are indicated with orange squares.}
    \label{fig:coreradius_coredensity_allclusters}
\end{figure}

We plot the core radius against the core density of all clusters in the final snapshot in Figure~\ref{fig:coreradius_coredensity_allclusters}. 
Here, we see that the `cloud' and `strong' simulations produce the largest number of star clusters, while the distributions of core radii and core densities are similar.
This suggests that the internal structure and densities of similar size clusters are similar regardless of the initial simulation, which is unsurprising as these are presumably driven mostly by N-body dynamics.
We also see that the most massive clusters exhibit the highest core densities. 

\begin{figure}
    \centering
    \includegraphics[width=0.9\columnwidth]{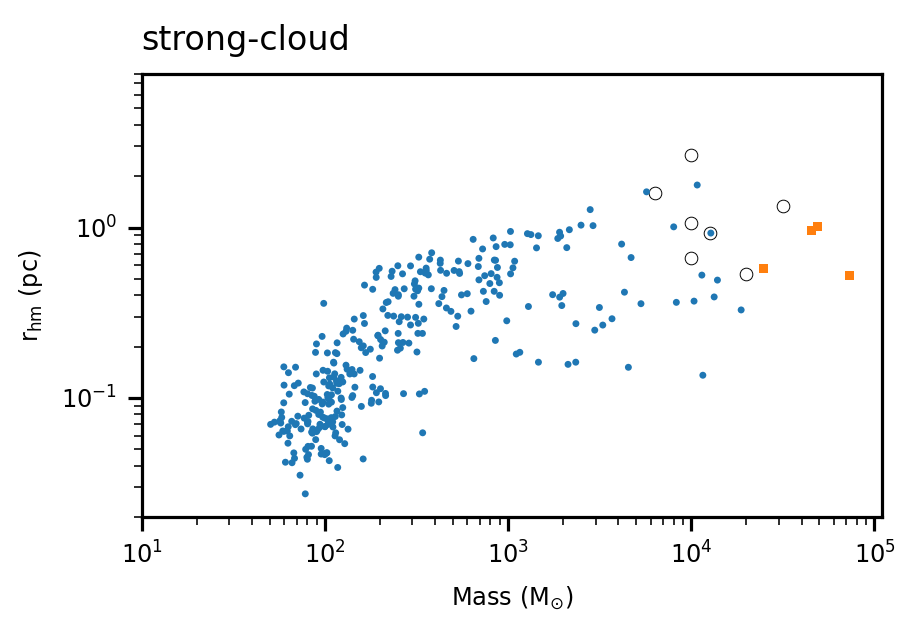}
    \includegraphics[width=0.9\columnwidth]{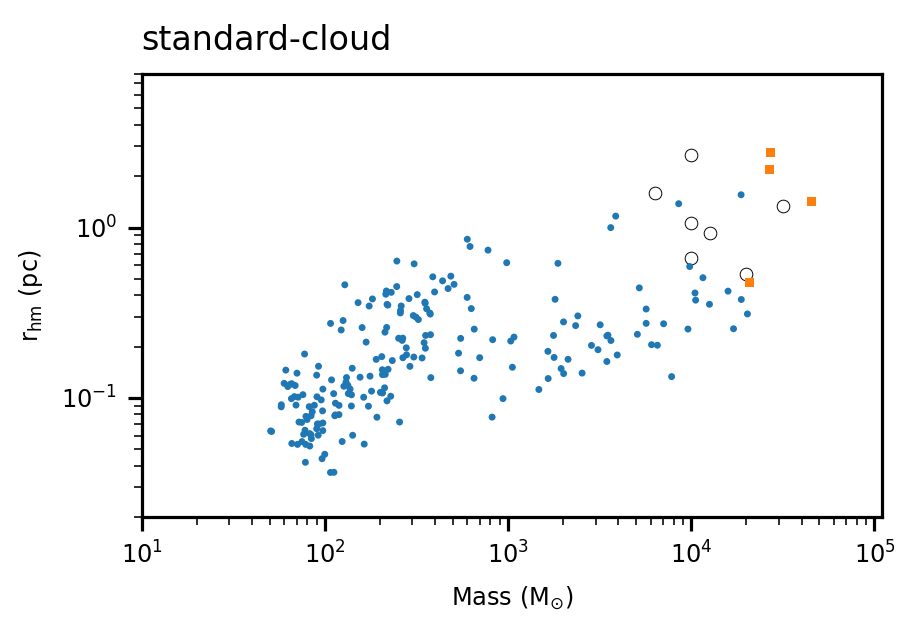}
    \includegraphics[width=0.9\columnwidth]{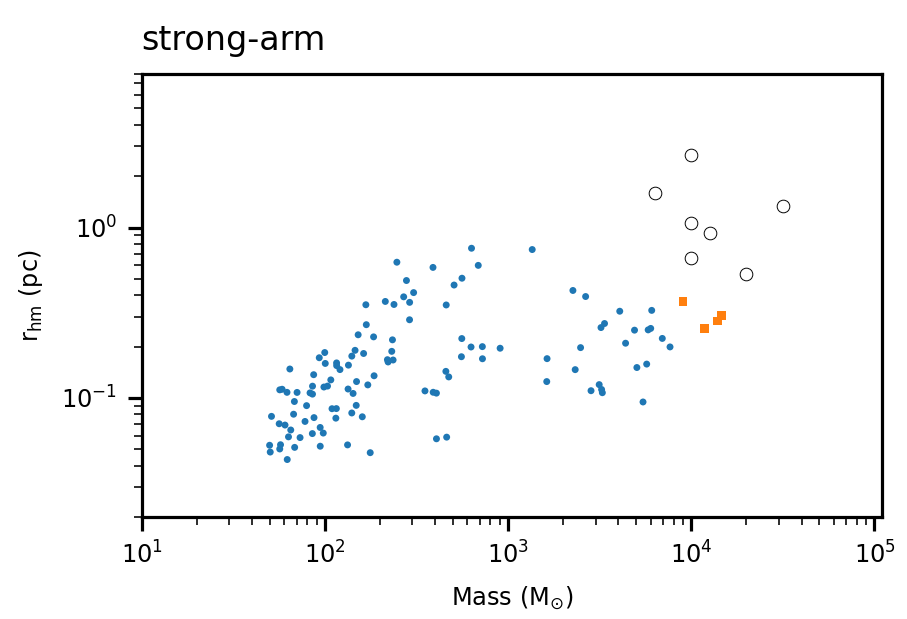}
    \includegraphics[width=0.9\columnwidth]{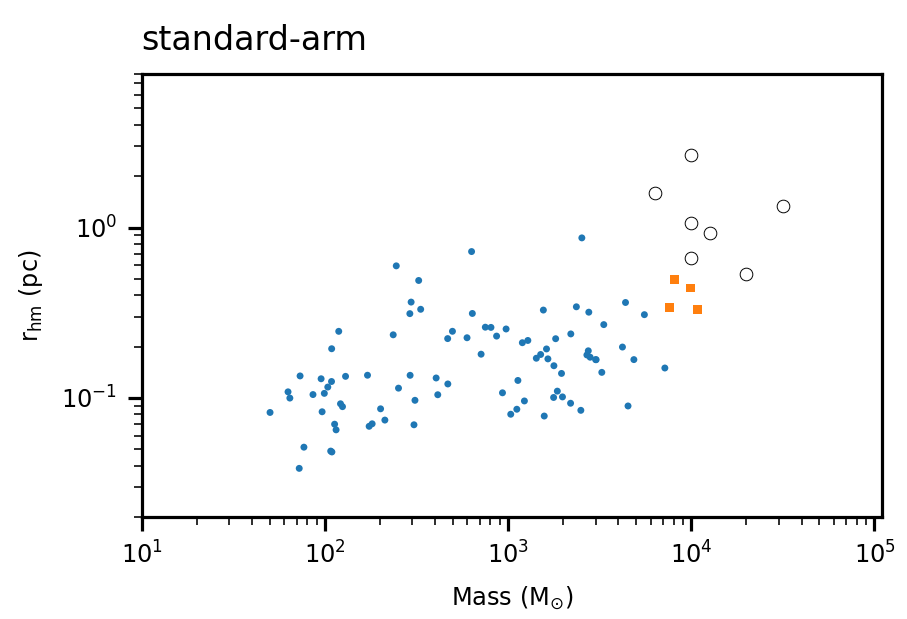}
    \caption{Mass versus half-mass radius for clusters at 3.0 Myr. The most massive clusters (orange squares) have properties comparable to the similarly aged observed young massive clusters. The lowest mass clusters ($\lesssim 200$ M$_{\odot}$) have comparable masses, but slightly smaller radii than those in \citet{ladalada}, however we caution that these clusters may be dominated by very recently inserted stars which have not yet evolved far from their initial conditions.}
    \label{fig:mass_halfmassradius}
\end{figure}

Finally, in Fig.~\ref{fig:mass_halfmassradius} we plot the cluster mass versus the half-mass radii, again at 3 Myr.
Generally there is not a strong dependence of radius versus mass, as seen in observations \citep{pzmg10}.
We also show the young massive clusters \citet[][and references therein]{pzmg10} plotted for comparison as open circles, which are similarly aged to our clusters ($<= 3.5$ Myr).
We see again that the `cloud' regions produce more massive star clusters than the `arm' regions.
When comparing our simulated clusters to the observed ones, we find that the more massive of our clusters have similar masses and half-mass radii.
\citet{pfalzner13} previously noted that observed young massive clusters exhibit comparable radii. 
For the lower mass clusters, there is a clearer increase of mass versus radius, roughly in line with the observed correlation between cluster masses and radii \citep{adams06,pzmg10,pfalzner16}. 
We do not show open clusters from \citet{pzmg10}, as these include clusters which are older than those in our simulations.
Open clusters tend to have larger radii for comparable masses, which could suggest that clusters expand over time, and move from the lower left or middle of our figure to the upper left, as found in simulations of lower mass clusters or associations by \citet{pfalzner13}.

\section{Conclusions}
\label{conclusions}
We have simulated the formation of star clusters in two different sections of two spiral galaxies, using the \Ekster{} method that combines high-precision \nbody{} dynamics with \ac{sph} and stellar evolution.
By comparing two \ac{gmc} regions and two spiral arm regions from two different galaxies, we find that the massive and strongly bound \ac{gmc}s are able to form larger star clusters in a shorter time compared to typical molecular clouds.
This is independent of the galaxy scale simulation from which the clouds are extracted, though such \ac{gmc}s are much more commonplace in our galaxy model with a stronger spiral potential.
These results agree with the hypothesis in \citet{dobbs20} that massive clusters (and massive \ac{gmc}s, \citep{dobbs11} form in more strongly converging flows.
This could potentially explain why star formation, and the properties of molecular clouds, appear to be more effected by spiral arms in for example M51, compared to our own Milky Way Galaxy \citep[Colombo et al., submitted]{colombo14}.

We also find that clusters partially grow by merging with other (proto-)clusters. This agrees with previous work \citep{sb17}, although here we are fully resolving the mergers in our models.
These mergers produce clear peaks in the velocity dispersion of the stars.
Again, this is more commonplace during the formation of massive clusters formed in massive \ac{gmc}s compared to lower mass clusters formed in lower mass clouds.
We compare the properties of our clusters with observed clusters.
We find that our more massive clusters have similar properties to observed young massive clusters, including a fairly constant mass radius relation.
Smaller clusters do show an increase in mass with radius, but tend to be lower radii compared to typical observed open clusters, possibly because these systems are generally younger.

\section*{Acknowledgements}

We thank the anonymous referee for some useful suggestions including further clarifying the dependence of the cloud properties on the original galaxy models.
We are grateful to Daniel~Price for his assistance with the {\tt Phantom} code, and to Long~Wang and Masaki~Iwasawa for their help with {\tt PeTar} and {\tt Pentacle}.
It is a pleasure to also acknowledge Inti~Pelupessy, for his continuing help with AMUSE.
Calculations for this paper were performed on the ISCA High Performance Computing Service at the University of Exeter, and used the DiRAC DIaL system, operated by the University of Leicester IT Services, which forms part of the STFC DiRAC HPC Facility (www.dirac.ac.uk).
The equipment was funded by BEIS capital funding via STFC capital grants ST/K000373/1 and ST/R002363/1 and STFC DiRAC Operations grant ST/K001014/1. 
DiRAC is part of the National E-Infrastructure. 
SR acknowledges funding from STFC Consolidated Grant ST/R000395/1.
CLD acknowledges funding from the European Research Council for the Horizon 2020 ERC consolidator grant project ICYBOB, grant number 818940.
Apart from software specifically mentioned above, the simulation and analysis in this article additionally would like to acknowledge the use of the following open-source software:
numpy \citep{numpy},
mpi4py \citep{mpi4py,mpi4py311},
matplotlib \citep{matplotlib,matplotlib332}.

\section*{Data availability}
The data underlying this article will be shared on reasonable request to the corresponding author.

\section*{Contributions}
The authors contributed to this article specifically in the following ways.
SR: concept, developing \Ekster, running and analysing the simulations.
CD: concept, initial galaxy simulations.
TB: initial galaxy simulations, zoom-in initial conditions.
KYL: star formation from sinks, co-developing \Ekster.
JW: assistance with and developing of {\tt Phantom}.



\bibliographystyle{mnras}
\bibliography{article}

\begin{thebibliography}{}
\makeatletter
\relax
\def\mn@urlcharsother{\let\do\@makeother \do\$\do\&\do\#\do\^\do\_\do\%\do\~}
\def\mn@doi{\begingroup\mn@urlcharsother \@ifnextchar [ {\mn@doi@}
  {\mn@doi@[]}}
\def\mn@doi@[#1]#2{\def\@tempa{#1}\ifx\@tempa\@empty \href
  {http://dx.doi.org/#2} {doi:#2}\else \href {http://dx.doi.org/#2} {#1}\fi
  \endgroup}
\def\mn@eprint#1#2{\mn@eprint@#1:#2::\@nil}
\def\mn@eprint@arXiv#1{\href {http://arxiv.org/abs/#1} {{\tt arXiv:#1}}}
\def\mn@eprint@dblp#1{\href {http://dblp.uni-trier.de/rec/bibtex/#1.xml}
  {dblp:#1}}
\def\mn@eprint@#1:#2:#3:#4\@nil{\def\@tempa {#1}\def\@tempb {#2}\def\@tempc
  {#3}\ifx \@tempc \@empty \let \@tempc \@tempb \let \@tempb \@tempa \fi \ifx
  \@tempb \@empty \def\@tempb {arXiv}\fi \@ifundefined
  {mn@eprint@\@tempb}{\@tempb:\@tempc}{\expandafter \expandafter \csname
  mn@eprint@\@tempb\endcsname \expandafter{\@tempc}}}

\bibitem[\protect\citeauthoryear{{Aarseth}}{{Aarseth}}{1974}]{aarseth74}
{Aarseth} S.~J.,  1974, \aap, \href
  {https://ui.adsabs.harvard.edu/abs/1974A&A....35..237A} {35, 237}

\bibitem[\protect\citeauthoryear{{Adams}, {Proszkow}, {Fatuzzo}  \&
  {Myers}}{{Adams} et~al.}{2006}]{adams06}
{Adams} F.~C.,  {Proszkow} E.~M.,  {Fatuzzo} M.,   {Myers} P.~C.,  2006,
  \mn@doi [\apj] {10.1086/500393}, \href
  {https://ui.adsabs.harvard.edu/abs/2006ApJ...641..504A} {641, 504}

\bibitem[\protect\citeauthoryear{{Alig}, {Hammer}, {Borodatchenkova}, {Dobbs}
  \& {Burkert}}{{Alig} et~al.}{2018}]{alig18}
{Alig} C.,  {Hammer} S.,  {Borodatchenkova} N.,  {Dobbs} C.~L.,   {Burkert} A.,
   2018, \mn@doi [\apjl] {10.3847/2041-8213/aaf1cb}, \href
  {https://ui.adsabs.harvard.edu/abs/2018ApJ...869L...2A} {869, L2}

\bibitem[\protect\citeauthoryear{{Allison}, {Goodwin}, {Parker}, {Portegies
  Zwart}  \& {de Grijs}}{{Allison} et~al.}{2010}]{allison10}
{Allison} R.~J.,  {Goodwin} S.~P.,  {Parker} R.~J.,  {Portegies Zwart} S.~F.,
  {de Grijs} R.,  2010, \mn@doi [\mnras] {10.1111/j.1365-2966.2010.16939.x},
  \href {https://ui.adsabs.harvard.edu/abs/2010MNRAS.407.1098A} {407, 1098}

\bibitem[\protect\citeauthoryear{{Banerjee} \& {Kroupa}}{{Banerjee} \&
  {Kroupa}}{2013}]{bk13}
{Banerjee} S.,  {Kroupa} P.,  2013, \mn@doi [\apj]
  {10.1088/0004-637X/764/1/29}, \href
  {https://ui.adsabs.harvard.edu/abs/2013ApJ...764...29B} {764, 29}

\bibitem[\protect\citeauthoryear{{Barnes} \& {Hut}}{{Barnes} \&
  {Hut}}{1986}]{barneshut}
{Barnes} J.,  {Hut} P.,  1986, \mn@doi [\nat] {10.1038/324446a0}, \href
  {https://ui.adsabs.harvard.edu/abs/1986Natur.324..446B} {324, 446}

\bibitem[\protect\citeauthoryear{{Bastian} \& {Goodwin}}{{Bastian} \&
  {Goodwin}}{2006}]{bg06}
{Bastian} N.,  {Goodwin} S.~P.,  2006, \mn@doi [\mnras]
  {10.1111/j.1745-3933.2006.00162.x}, \href
  {https://ui.adsabs.harvard.edu/abs/2006MNRAS.369L...9B} {369, L9}

\bibitem[\protect\citeauthoryear{{Bastian}, {Gieles}, {Efremov}  \&
  {Lamers}}{{Bastian} et~al.}{2005}]{bastian05}
{Bastian} N.,  {Gieles} M.,  {Efremov} Y.~N.,   {Lamers} H.~J.~G.~L.~M.,  2005,
  \mn@doi [\aap] {10.1051/0004-6361:20053165}, \href
  {https://ui.adsabs.harvard.edu/abs/2005A&A...443...79B} {443, 79}

\bibitem[\protect\citeauthoryear{{Bate}}{{Bate}}{2012}]{bate12}
{Bate} M.~R.,  2012, \mn@doi [\mnras] {10.1111/j.1365-2966.2011.19955.x}, \href
  {https://ui.adsabs.harvard.edu/abs/2012MNRAS.419.3115B} {419, 3115}

\bibitem[\protect\citeauthoryear{{Bate}, {Bonnell}  \& {Price}}{{Bate}
  et~al.}{1995}]{bate95}
{Bate} M.~R.,  {Bonnell} I.~A.,   {Price} N.~M.,  1995, \mn@doi [\mnras]
  {10.1093/mnras/277.2.362}, \href
  {https://ui.adsabs.harvard.edu/abs/1995MNRAS.277..362B} {277, 362}

\bibitem[\protect\citeauthoryear{{Bate}, {Bonnell}  \& {Bromm}}{{Bate}
  et~al.}{2003}]{bate2003}
{Bate} M.~R.,  {Bonnell} I.~A.,   {Bromm} V.,  2003, \mn@doi [\mnras]
  {10.1046/j.1365-8711.2003.06210.x}, \href
  {https://ui.adsabs.harvard.edu/abs/2003MNRAS.339..577B} {339, 577}

\bibitem[\protect\citeauthoryear{{Baumgardt} \& {Kroupa}}{{Baumgardt} \&
  {Kroupa}}{2007}]{bk07}
{Baumgardt} H.,  {Kroupa} P.,  2007, \mn@doi [\mnras]
  {10.1111/j.1365-2966.2007.12209.x}, \href
  {https://ui.adsabs.harvard.edu/abs/2007MNRAS.380.1589B} {380, 1589}

\bibitem[\protect\citeauthoryear{{Bending}, {Dobbs}  \& {Bate}}{{Bending}
  et~al.}{2020}]{bending20}
{Bending} T. J.~R.,  {Dobbs} C.~L.,   {Bate} M.~R.,  2020, \mn@doi [\mnras]
  {10.1093/mnras/staa1293}, \href
  {https://ui.adsabs.harvard.edu/abs/2020MNRAS.495.1672B} {495, 1672}

\bibitem[\protect\citeauthoryear{{Binney} \& {Tremaine}}{{Binney} \&
  {Tremaine}}{2008}]{galacticdynamics}
{Binney} J.,  {Tremaine} S.,  2008, {Galactic Dynamics}.
{Princeton University Press}

\bibitem[\protect\citeauthoryear{{Bonnell}, {Bate}  \& {Vine}}{{Bonnell}
  et~al.}{2003}]{bonnell03}
{Bonnell} I.~A.,  {Bate} M.~R.,   {Vine} S.~G.,  2003, \mn@doi [\mnras]
  {10.1046/j.1365-8711.2003.06687.x}, \href
  {https://ui.adsabs.harvard.edu/abs/2003MNRAS.343..413B} {343, 413}

\bibitem[\protect\citeauthoryear{{Buckner} et~al.,}{{Buckner}
  et~al.}{2019}]{buckner19}
{Buckner} A. S.~M.,  et~al., 2019, \mn@doi [\aap]
  {10.1051/0004-6361/201832936}, \href
  {https://ui.adsabs.harvard.edu/abs/2019A&A...622A.184B} {622, A184}

\bibitem[\protect\citeauthoryear{Caswell et~al.,}{Caswell
  et~al.}{2020}]{matplotlib332}
Caswell T.~A.,  et~al., 2020, matplotlib/matplotlib: REL: v3.3.2,
  \mn@doi{10.5281/zenodo.4030140}, \url
  {https://doi.org/10.5281/zenodo.4030140}

\bibitem[\protect\citeauthoryear{{Chen}, {Li}  \& {Vogelsberger}}{{Chen}
  et~al.}{2021}]{chen21}
{Chen} Y.,  {Li} H.,   {Vogelsberger} M.,  2021, \mn@doi [\mnras]
  {10.1093/mnras/stab491}, \href
  {https://ui.adsabs.harvard.edu/abs/2021MNRAS.502.6157C} {502, 6157}

\bibitem[\protect\citeauthoryear{{Colombo} et~al.,}{{Colombo}
  et~al.}{2014}]{colombo14}
{Colombo} D.,  et~al., 2014, \mn@doi [\apj] {10.1088/0004-637X/784/1/3}, \href
  {https://ui.adsabs.harvard.edu/abs/2014ApJ...784....3C} {784, 3}

\bibitem[\protect\citeauthoryear{{Cox} \& {G{\'o}mez}}{{Cox} \&
  {G{\'o}mez}}{2002}]{2002ApJS..142..261C}
{Cox} D.~P.,  {G{\'o}mez} G.~C.,  2002, \apjs, \href
  {https://ui.adsabs.harvard.edu/abs/2002ApJS..142..261C} {142, 261}

\bibitem[\protect\citeauthoryear{{Daffern-Powell} \& {Parker}}{{Daffern-Powell}
  \& {Parker}}{2020}]{dpp20}
{Daffern-Powell} E.~C.,  {Parker} R.~J.,  2020, \mn@doi [\mnras]
  {10.1093/mnras/staa575}, \href
  {https://ui.adsabs.harvard.edu/abs/2020MNRAS.493.4925D} {493, 4925}

\bibitem[\protect\citeauthoryear{Dalcin \& Fang}{Dalcin \&
  Fang}{2021a}]{mpi4py311}
Dalcin L.,  Fang Y.-L.~L.,  2021a, MPI for Python,
  \mn@doi{10.5281/zenodo.5229640}, \url
  {https://doi.org/10.5281/zenodo.5229640}

\bibitem[\protect\citeauthoryear{Dalcin \& Fang}{Dalcin \&
  Fang}{2021b}]{mpi4py}
Dalcin L.,  Fang Y.-L.~L.,  2021b, \mn@doi [Computing in Science Engineering]
  {10.1109/MCSE.2021.3083216}, 23, 47

\bibitem[\protect\citeauthoryear{{D{\'\i}az-Garc{\'\i}a}, {Moyano},
  {Comer{\'o}n}, {Knapen}, {Salo}  \& {Bouquin}}{{D{\'\i}az-Garc{\'\i}a}
  et~al.}{2020}]{diazgarcia20}
{D{\'\i}az-Garc{\'\i}a} S.,  {Moyano} F.~D.,  {Comer{\'o}n} S.,  {Knapen}
  J.~H.,  {Salo} H.,   {Bouquin} A.~Y.~K.,  2020, \mn@doi [\aap]
  {10.1051/0004-6361/202039162}, \href
  {https://ui.adsabs.harvard.edu/abs/2020A&A...644A..38D} {644, A38}

\bibitem[\protect\citeauthoryear{{Dobbs} \& {Pringle}}{{Dobbs} \&
  {Pringle}}{2013}]{dobbs13}
{Dobbs} C.~L.,  {Pringle} J.~E.,  2013, \mnras, \href
  {https://ui.adsabs.harvard.edu/abs/2013MNRAS.432..653D} {432, 653}

\bibitem[\protect\citeauthoryear{{Dobbs}, {Burkert}  \& {Pringle}}{{Dobbs}
  et~al.}{2011}]{dobbs11}
{Dobbs} C.~L.,  {Burkert} A.,   {Pringle} J.~E.,  2011, \mn@doi [\mnras]
  {10.1111/j.1365-2966.2011.19346.x}, \href
  {https://ui.adsabs.harvard.edu/abs/2011MNRAS.417.1318D} {417, 1318}

\bibitem[\protect\citeauthoryear{{Dobbs}, {Pringle}  \&
  {Duarte-Cabral}}{{Dobbs} et~al.}{2015}]{dobbs15}
{Dobbs} C.~L.,  {Pringle} J.~E.,   {Duarte-Cabral} A.,  2015, \mn@doi [\mnras]
  {10.1093/mnras/stu2319}, \href
  {https://ui.adsabs.harvard.edu/abs/2015MNRAS.446.3608D} {446, 3608}

\bibitem[\protect\citeauthoryear{{Dobbs}, {Liow}  \& {Rieder}}{{Dobbs}
  et~al.}{2020}]{dobbs20}
{Dobbs} C.~L.,  {Liow} K.~Y.,   {Rieder} S.,  2020, \mn@doi [\mnras]
  {10.1093/mnrasl/slaa072}, \href
  {https://ui.adsabs.harvard.edu/abs/2020MNRAS.496L...1D} {496, L1}

\bibitem[\protect\citeauthoryear{{Dobbs}, {Bending}, {Pettitt}  \&
  {Bate}}{{Dobbs} et~al.}{2021}]{dobbs21}
{Dobbs} C.~L.,  {Bending} T.~J.~R.,  {Pettitt} A.~R.,   {Bate} M.~R.,  2021,
  \mn@doi [\mnras] {10.1093/mnras/stab3036}, \href
  {https://ui.adsabs.harvard.edu/abs/2021MNRAS.tmp.2789D} {}

\bibitem[\protect\citeauthoryear{{Eden}, {Moore}, {Plume}  \& {Morgan}}{{Eden}
  et~al.}{2012}]{eden12}
{Eden} D.~J.,  {Moore} T.~J.~T.,  {Plume} R.,   {Morgan} L.~K.,  2012, \mn@doi
  [\mnras] {10.1111/j.1365-2966.2012.20840.x}, \href
  {https://ui.adsabs.harvard.edu/abs/2012MNRAS.422.3178E} {422, 3178}

\bibitem[\protect\citeauthoryear{{Eden}, {Moore}, {Morgan}, {Thompson}  \&
  {Urquhart}}{{Eden} et~al.}{2013}]{eden13}
{Eden} D.~J.,  {Moore} T.~J.~T.,  {Morgan} L.~K.,  {Thompson} M.~A.,
  {Urquhart} J.~S.,  2013, \mn@doi [\mnras] {10.1093/mnras/stt279}, \href
  {https://ui.adsabs.harvard.edu/abs/2013MNRAS.431.1587E} {431, 1587}

\bibitem[\protect\citeauthoryear{{Eden}, {Moore}, {Urquhart}, {Elia}, {Plume},
  {Rigby}  \& {Thompson}}{{Eden} et~al.}{2015}]{eden15}
{Eden} D.~J.,  {Moore} T.~J.~T.,  {Urquhart} J.~S.,  {Elia} D.,  {Plume} R.,
  {Rigby} A.~J.,   {Thompson} M.~A.,  2015, \mn@doi [\mnras]
  {10.1093/mnras/stv1323}, \href
  {https://ui.adsabs.harvard.edu/abs/2015MNRAS.452..289E} {452, 289}

\bibitem[\protect\citeauthoryear{{Eisenstein} \& {Hut}}{{Eisenstein} \&
  {Hut}}{1998}]{hop}
{Eisenstein} D.~J.,  {Hut} P.,  1998, \mn@doi [\apj] {10.1086/305535}, \href
  {https://ui.adsabs.harvard.edu/abs/1998ApJ...498..137E} {498, 137}

\bibitem[\protect\citeauthoryear{{Elmegreen} \& {Elmegreen}}{{Elmegreen} \&
  {Elmegreen}}{1986}]{elmegreen86}
{Elmegreen} B.~G.,  {Elmegreen} D.~M.,  1986, \mn@doi [\apj] {10.1086/164795},
  \href {https://ui.adsabs.harvard.edu/abs/1986ApJ...311..554E} {311, 554}

\bibitem[\protect\citeauthoryear{{Emsellem}, {Renaud}, {Bournaud}, {Elmegreen},
  {Combes}  \& {Gabor}}{{Emsellem} et~al.}{2015}]{emsellem15}
{Emsellem} E.,  {Renaud} F.,  {Bournaud} F.,  {Elmegreen} B.,  {Combes} F.,
  {Gabor} J.~M.,  2015, \mn@doi [\mnras] {10.1093/mnras/stu2209}, \href
  {https://ui.adsabs.harvard.edu/abs/2015MNRAS.446.2468E} {446, 2468}

\bibitem[\protect\citeauthoryear{{Farias}, {Smith}, {Fellhauer}, {Goodwin},
  {Candlish}, {Bla{\~n}a}  \& {Dominguez}}{{Farias} et~al.}{2015}]{farias15}
{Farias} J.~P.,  {Smith} R.,  {Fellhauer} M.,  {Goodwin} S.,  {Candlish} G.~N.,
   {Bla{\~n}a} M.,   {Dominguez} R.,  2015, \mn@doi [\mnras]
  {10.1093/mnras/stv790}, \href
  {https://ui.adsabs.harvard.edu/abs/2015MNRAS.450.2451F} {450, 2451}

\bibitem[\protect\citeauthoryear{{Fujii} \& {Portegies Zwart}}{{Fujii} \&
  {Portegies Zwart}}{2015}]{fpz15}
{Fujii} M.~S.,  {Portegies Zwart} S.,  2015, \mn@doi [\mnras]
  {10.1093/mnras/stv293}, \href
  {https://ui.adsabs.harvard.edu/abs/2015MNRAS.449..726F} {449, 726}

\bibitem[\protect\citeauthoryear{{Fujii}, {Iwasawa}, {Funato}  \&
  {Makino}}{{Fujii} et~al.}{2007}]{2007PASJ...59.1095F}
{Fujii} M.,  {Iwasawa} M.,  {Funato} Y.,   {Makino} J.,  2007, \pasj, \href
  {https://ui.adsabs.harvard.edu/abs/2007PASJ...59.1095F} {59, 1095}

\bibitem[\protect\citeauthoryear{{Fujii}, {Saitoh}, {Wang}  \& {Hirai}}{{Fujii}
  et~al.}{2021}]{fujii21a}
{Fujii} M.~S.,  {Saitoh} T.~R.,  {Wang} L.,   {Hirai} Y.,  2021, \mn@doi
  [\pasj] {10.1093/pasj/psab037}, \href
  {https://ui.adsabs.harvard.edu/abs/2021PASJ...73.1057F} {73, 1057}

\bibitem[\protect\citeauthoryear{{Gafton} \& {Rosswog}}{{Gafton} \&
  {Rosswog}}{2011}]{gaftonrosswog11}
{Gafton} E.,  {Rosswog} S.,  2011, \mn@doi [\mnras]
  {10.1111/j.1365-2966.2011.19528.x}, \href
  {https://ui.adsabs.harvard.edu/abs/2011MNRAS.418..770G} {418, 770}

\bibitem[\protect\citeauthoryear{{Geyer} \& {Burkert}}{{Geyer} \&
  {Burkert}}{2001}]{gb01}
{Geyer} M.~P.,  {Burkert} A.,  2001, \mn@doi [\mnras]
  {10.1046/j.1365-8711.2001.04257.x}, \href
  {https://ui.adsabs.harvard.edu/abs/2001MNRAS.323..988G} {323, 988}

\bibitem[\protect\citeauthoryear{{Glover} \& {Mac Low}}{{Glover} \& {Mac
  Low}}{2007}]{glover07}
{Glover} S. C.~O.,  {Mac Low} M.-M.,  2007, \mn@doi [\apjs] {10.1086/512238},
  \href {https://ui.adsabs.harvard.edu/abs/2007ApJS..169..239G} {169, 239}

\bibitem[\protect\citeauthoryear{{Grudi{\'c}}, {Guszejnov}, {Hopkins},
  {Lamberts}, {Boylan-Kolchin}, {Murray}  \& {Schmitz}}{{Grudi{\'c}}
  et~al.}{2018}]{grudic18}
{Grudi{\'c}} M.~Y.,  {Guszejnov} D.,  {Hopkins} P.~F.,  {Lamberts} A.,
  {Boylan-Kolchin} M.,  {Murray} N.,   {Schmitz} D.,  2018, \mn@doi [\mnras]
  {10.1093/mnras/sty2303}, \href
  {https://ui.adsabs.harvard.edu/abs/2018MNRAS.481..688G} {481, 688}

\bibitem[\protect\citeauthoryear{Harris et~al.,}{Harris et~al.}{2020}]{numpy}
Harris C.~R.,  et~al., 2020, \mn@doi [Nature] {10.1038/s41586-020-2649-2}, 585,
  357

\bibitem[\protect\citeauthoryear{{Heggie} \& {Hut}}{{Heggie} \&
  {Hut}}{2003}]{heggiehut}
{Heggie} D.,  {Hut} P.,  2003, {The Gravitational Million-Body Problem: A
  Multidisciplinary Approach to Star Cluster Dynamics}

\bibitem[\protect\citeauthoryear{{Hirai}, {Fujii}  \& {Saitoh}}{{Hirai}
  et~al.}{2021}]{hirai21}
{Hirai} Y.,  {Fujii} M.~S.,   {Saitoh} T.~R.,  2021, \mn@doi [\pasj]
  {10.1093/pasj/psab038}, \href
  {https://ui.adsabs.harvard.edu/abs/2021PASJ...73.1036H} {73, 1036}

\bibitem[\protect\citeauthoryear{{Hirota} et~al.,}{{Hirota}
  et~al.}{2014}]{hirota14}
{Hirota} A.,  et~al., 2014, \mn@doi [\pasj] {10.1093/pasj/psu006}, \href
  {https://ui.adsabs.harvard.edu/abs/2014PASJ...66...46H} {66, 46}

\bibitem[\protect\citeauthoryear{{Howard}, {Pudritz}, {Sills}  \&
  {Harris}}{{Howard} et~al.}{2019}]{howard19}
{Howard} C.~S.,  {Pudritz} R.~E.,  {Sills} A.,   {Harris} W.~E.,  2019, \mn@doi
  [\mnras] {10.1093/mnras/stz924}, \href
  {https://ui.adsabs.harvard.edu/abs/2019MNRAS.486.1146H} {486, 1146}

\bibitem[\protect\citeauthoryear{{Hubber}, {Allison}, {Smith}  \&
  {Goodwin}}{{Hubber} et~al.}{2013}]{hubber13}
{Hubber} D.~A.,  {Allison} R.~J.,  {Smith} R.,   {Goodwin} S.~P.,  2013,
  \mn@doi [\mnras] {10.1093/mnras/sts694}, \href
  {https://ui.adsabs.harvard.edu/abs/2013MNRAS.430.1599H} {430, 1599}

\bibitem[\protect\citeauthoryear{{Hunter}}{{Hunter}}{2007}]{matplotlib}
{Hunter} J.~D.,  2007, \mn@doi [Computing in Science and Engineering]
  {10.1109/MCSE.2007.55}, \href
  {https://ui.adsabs.harvard.edu/abs/2007CSE.....9...90H} {9, 90}

\bibitem[\protect\citeauthoryear{{Inutsuka}, {Inoue}, {Iwasaki}  \&
  {Hosokawa}}{{Inutsuka} et~al.}{2015}]{inutsuka15}
{Inutsuka} S.-i.,  {Inoue} T.,  {Iwasaki} K.,   {Hosokawa} T.,  2015, \mn@doi
  [\aap] {10.1051/0004-6361/201425584}, \href
  {https://ui.adsabs.harvard.edu/abs/2015A&A...580A..49I} {580, A49}

\bibitem[\protect\citeauthoryear{{Iwasawa}, {Oshino}, {Fujii}  \&
  {Hori}}{{Iwasawa} et~al.}{2017}]{iwasawa17}
{Iwasawa} M.,  {Oshino} S.,  {Fujii} M.~S.,   {Hori} Y.,  2017, \mn@doi [\pasj]
  {10.1093/pasj/psx073}, \href
  {https://ui.adsabs.harvard.edu/abs/2017PASJ...69...81I} {69, 81}

\bibitem[\protect\citeauthoryear{{Kim}, {Kim}  \& {Ostriker}}{{Kim}
  et~al.}{2020}]{kimwt20}
{Kim} W.-T.,  {Kim} C.-G.,   {Ostriker} E.~C.,  2020, \mn@doi [\apj]
  {10.3847/1538-4357/ab9b87}, \href
  {https://ui.adsabs.harvard.edu/abs/2020ApJ...898...35K} {898, 35}

\bibitem[\protect\citeauthoryear{{Kim} et~al.,}{{Kim} et~al.}{2021}]{kimj21}
{Kim} J.,  et~al., 2021, \mn@doi [\mnras] {10.1093/mnras/stab878}, \href
  {https://ui.adsabs.harvard.edu/abs/2021MNRAS.504..487K} {504, 487}

\bibitem[\protect\citeauthoryear{{King}}{{King}}{1966}]{king66}
{King} I.~R.,  1966, \mn@doi [\aj] {10.1086/109857}, \href
  {https://ui.adsabs.harvard.edu/abs/1966AJ.....71...64K} {71, 64}

\bibitem[\protect\citeauthoryear{{Kroupa}}{{Kroupa}}{2001}]{kroupa}
{Kroupa} P.,  2001, \mnras, \href
  {https://ui.adsabs.harvard.edu/abs/2001MNRAS.322..231K} {322, 231}

\bibitem[\protect\citeauthoryear{{Lada} \& {Lada}}{{Lada} \&
  {Lada}}{2003}]{ladalada}
{Lada} C.~J.,  {Lada} E.~A.,  2003, \mn@doi [\araa]
  {10.1146/annurev.astro.41.011802.094844}, \href
  {https://ui.adsabs.harvard.edu/abs/2003ARA&A..41...57L} {41, 57}

\bibitem[\protect\citeauthoryear{{Liow} \& {Dobbs}}{{Liow} \&
  {Dobbs}}{2020}]{Liow20}
{Liow} K.~Y.,  {Dobbs} C.~L.,  2020, \mn@doi [\mnras] {10.1093/mnras/staa2857},
  \href {https://ui.adsabs.harvard.edu/abs/2020MNRAS.499.1099L} {499, 1099}

\bibitem[\protect\citeauthoryear{{Maeda}, {Ohta}, {Fujimoto}, {Habe}  \&
  {Ushio}}{{Maeda} et~al.}{2020}]{maeda20}
{Maeda} F.,  {Ohta} K.,  {Fujimoto} Y.,  {Habe} A.,   {Ushio} K.,  2020,
  \mn@doi [\mnras] {10.1093/mnras/staa1296}, \href
  {https://ui.adsabs.harvard.edu/abs/2020MNRAS.495.3840M} {495, 3840}

\bibitem[\protect\citeauthoryear{{Moeckel} \& {Bate}}{{Moeckel} \&
  {Bate}}{2010}]{mb10}
{Moeckel} N.,  {Bate} M.~R.,  2010, \mn@doi [\mnras]
  {10.1111/j.1365-2966.2010.16347.x}, \href
  {https://ui.adsabs.harvard.edu/abs/2010MNRAS.404..721M} {404, 721}

\bibitem[\protect\citeauthoryear{{Motte} et~al.,}{{Motte}
  et~al.}{2014}]{motte14}
{Motte} F.,  et~al., 2014, \mn@doi [\aap] {10.1051/0004-6361/201323001}, \href
  {https://ui.adsabs.harvard.edu/abs/2014A&A...571A..32M} {571, A32}

\bibitem[\protect\citeauthoryear{{Ochsendorf}, {Zinnecker}, {Nayak}, {Bally},
  {Meixner}, {Jones}, {Indebetouw}  \& {Rahman}}{{Ochsendorf}
  et~al.}{2017}]{oschendorf17}
{Ochsendorf} B.~B.,  {Zinnecker} H.,  {Nayak} O.,  {Bally} J.,  {Meixner} M.,
  {Jones} O.~C.,  {Indebetouw} R.,   {Rahman} M.,  2017, \mn@doi [Nature
  Astronomy] {10.1038/s41550-017-0268-0}, \href
  {https://ui.adsabs.harvard.edu/abs/2017NatAs...1..784O} {1, 784}

\bibitem[\protect\citeauthoryear{{Pecaut} \& {Mamajek}}{{Pecaut} \&
  {Mamajek}}{2016}]{pm16}
{Pecaut} M.~J.,  {Mamajek} E.~E.,  2016, \mn@doi [\mnras]
  {10.1093/mnras/stw1300}, \href
  {https://ui.adsabs.harvard.edu/abs/2016MNRAS.461..794P} {461, 794}

\bibitem[\protect\citeauthoryear{{Pelupessy}}{{Pelupessy}}{2005}]{ficode}
{Pelupessy} F.~I.,  2005, PhD thesis, Leiden Observatory, Leiden University,
  P.O. Box 9513, 2300 RA Leiden, The Netherlands

\bibitem[\protect\citeauthoryear{{Pelupessy}, {van Elteren}, {de Vries},
  {McMillan}, {Drost}  \& {Portegies Zwart}}{{Pelupessy}
  et~al.}{2013}]{2013A&A...557A..84P}
{Pelupessy} F.~I.,  {van Elteren} A.,  {de Vries} N.,  {McMillan} S.~L.~W.,
  {Drost} N.,   {Portegies Zwart} S.~F.,  2013, \aap, \href
  {https://ui.adsabs.harvard.edu/abs/2013A&A...557A..84P} {557, A84}

\bibitem[\protect\citeauthoryear{{Pettitt}, {Tasker}, {Wadsley}, {Keller}  \&
  {Benincasa}}{{Pettitt} et~al.}{2017}]{pettitt17}
{Pettitt} A.~R.,  {Tasker} E.~J.,  {Wadsley} J.~W.,  {Keller} B.~W.,
  {Benincasa} S.~M.,  2017, \mn@doi [\mnras] {10.1093/mnras/stx736}, \href
  {https://ui.adsabs.harvard.edu/abs/2017MNRAS.468.4189P} {468, 4189}

\bibitem[\protect\citeauthoryear{{Pettitt}, {Egusa}, {Dobbs}, {Tasker},
  {Fujimoto}  \& {Habe}}{{Pettitt} et~al.}{2018}]{pettitt18}
{Pettitt} A.~R.,  {Egusa} F.,  {Dobbs} C.~L.,  {Tasker} E.~J.,  {Fujimoto} Y.,
   {Habe} A.,  2018, \mn@doi [\mnras] {10.1093/mnras/sty2040}, \href
  {https://ui.adsabs.harvard.edu/abs/2018MNRAS.480.3356P} {480, 3356}

\bibitem[\protect\citeauthoryear{{Pfalzner} \& {Kaczmarek}}{{Pfalzner} \&
  {Kaczmarek}}{2013}]{pfalzner13}
{Pfalzner} S.,  {Kaczmarek} T.,  2013, \mn@doi [\aap]
  {10.1051/0004-6361/201322134}, \href
  {https://ui.adsabs.harvard.edu/abs/2013A&A...559A..38P} {559, A38}

\bibitem[\protect\citeauthoryear{{Pfalzner}, {Kirk}, {Sills}, {Urquhart},
  {Kauffmann}, {Kuhn}, {Bhandare}  \& {Menten}}{{Pfalzner}
  et~al.}{2016}]{pfalzner16}
{Pfalzner} S.,  {Kirk} H.,  {Sills} A.,  {Urquhart} J.~S.,  {Kauffmann} J.,
  {Kuhn} M.~A.,  {Bhandare} A.,   {Menten} K.~M.,  2016, \mn@doi [\aap]
  {10.1051/0004-6361/201527449}, \href
  {https://ui.adsabs.harvard.edu/abs/2016A&A...586A..68P} {586, A68}

\bibitem[\protect\citeauthoryear{{Plummer}}{{Plummer}}{1911}]{plummer}
{Plummer} H.~C.,  1911, \mn@doi [\mnras] {10.1093/mnras/71.5.460}, \href
  {https://ui.adsabs.harvard.edu/abs/1911MNRAS..71..460P} {71, 460}

\bibitem[\protect\citeauthoryear{Portegies~Zwart \& McMillan}{Portegies~Zwart
  \& McMillan}{2018}]{10.1088/978-0-7503-1320-9}
Portegies~Zwart S.,  McMillan S.,  2018, {Astrophysical Recipes; The art of
  AMUSE}.
2514-3433, IOP Publishing, \url {http://dx.doi.org/10.1088/978-0-7503-1320-9}

\bibitem[\protect\citeauthoryear{{Portegies Zwart} \& {Verbunt}}{{Portegies
  Zwart} \& {Verbunt}}{1996}]{seba1}
{Portegies Zwart} S.~F.,  {Verbunt} F.,  1996, \aap, \href
  {https://ui.adsabs.harvard.edu/abs/1996A&A...309..179P} {309, 179}

\bibitem[\protect\citeauthoryear{{Portegies Zwart} et~al.,}{{Portegies Zwart}
  et~al.}{2009}]{2009NewA...14..369P}
{Portegies Zwart} S.,  et~al., 2009, \na, \href
  {https://ui.adsabs.harvard.edu/abs/2009NewA...14..369P} {14, 369}

\bibitem[\protect\citeauthoryear{{Portegies Zwart}, {McMillan}  \&
  {Gieles}}{{Portegies Zwart} et~al.}{2010}]{pzmg10}
{Portegies Zwart} S.~F.,  {McMillan} S. L.~W.,   {Gieles} M.,  2010, \araa,
  \href {https://ui.adsabs.harvard.edu/abs/2010ARA&A..48..431P} {48, 431}

\bibitem[\protect\citeauthoryear{{Portegies Zwart}, {Pelupessy},
  {Mart{\'\i}nez-Barbosa}, {van Elteren}  \& {McMillan}}{{Portegies Zwart}
  et~al.}{2020}]{bridge}
{Portegies Zwart} S.,  {Pelupessy} I.,  {Mart{\'\i}nez-Barbosa} C.,  {van
  Elteren} A.,   {McMillan} S.,  2020, \mn@doi [Communications in Nonlinear
  Science and Numerical Simulations] {10.1016/j.cnsns.2020.105240}, \href
  {https://ui.adsabs.harvard.edu/abs/2020CNSNS..8505240P} {85, 105240}

\bibitem[\protect\citeauthoryear{{Preibisch} et~al.,}{{Preibisch}
  et~al.}{2011}]{preibisch11}
{Preibisch} T.,  et~al., 2011, \mn@doi [\apjs] {10.1088/0067-0049/194/1/10},
  \href {https://ui.adsabs.harvard.edu/abs/2011ApJS..194...10P} {194, 10}

\bibitem[\protect\citeauthoryear{{Price} et~al.,}{{Price}
  et~al.}{2018}]{phantom}
{Price} D.~J.,  et~al., 2018, \pasa, \href
  {https://ui.adsabs.harvard.edu/abs/2018PASA...35...31P} {35, e031}

\bibitem[\protect\citeauthoryear{{Renaud} et~al.,}{{Renaud}
  et~al.}{2015}]{renaud15}
{Renaud} F.,  et~al., 2015, \mn@doi [\mnras] {10.1093/mnras/stv2223}, \href
  {https://ui.adsabs.harvard.edu/abs/2015MNRAS.454.3299R} {454, 3299}

\bibitem[\protect\citeauthoryear{Rieder \& Liow}{Rieder \& Liow}{2021}]{ekster}
Rieder S.,  Liow K.~Y.,  2021, rieder/ekster:, \mn@doi{10.5281/zenodo.5520944},
  \url {https://doi.org/10.5281/zenodo.5520944}

\bibitem[\protect\citeauthoryear{{Rieder}, {Ishiyama}, {Langelaan}, {Makino},
  {McMillan}  \& {Portegies Zwart}}{{Rieder} et~al.}{2013}]{rieder13}
{Rieder} S.,  {Ishiyama} T.,  {Langelaan} P.,  {Makino} J.,  {McMillan} S.
  L.~W.,   {Portegies Zwart} S.,  2013, \mnras, \href
  {https://ui.adsabs.harvard.edu/abs/2013MNRAS.436.3695R} {436, 3695}

\bibitem[\protect\citeauthoryear{{Scheepmaker}, {Lamers}, {Anders}  \&
  {Larsen}}{{Scheepmaker} et~al.}{2009}]{scheepmaker09}
{Scheepmaker} R.~A.,  {Lamers} H.~J.~G.~L.~M.,  {Anders} P.,   {Larsen} S.~S.,
  2009, \mn@doi [\aap] {10.1051/0004-6361:200811068}, \href
  {https://ui.adsabs.harvard.edu/abs/2009A&A...494...81S} {494, 81}

\bibitem[\protect\citeauthoryear{{Sheth}, {Vogel}, {Regan}, {Thornley}  \&
  {Teuben}}{{Sheth} et~al.}{2005}]{sheth05}
{Sheth} K.,  {Vogel} S.~N.,  {Regan} M.~W.,  {Thornley} M.~D.,   {Teuben}
  P.~J.,  2005, \mn@doi [\apj] {10.1086/432409}, \href
  {https://ui.adsabs.harvard.edu/abs/2005ApJ...632..217S} {632, 217}

\bibitem[\protect\citeauthoryear{{Shukirgaliyev}, {Parmentier}, {Just}  \&
  {Berczik}}{{Shukirgaliyev} et~al.}{2018}]{shukirgaliyev18}
{Shukirgaliyev} B.,  {Parmentier} G.,  {Just} A.,   {Berczik} P.,  2018,
  \mn@doi [\apj] {10.3847/1538-4357/aad3bf}, \href
  {https://ui.adsabs.harvard.edu/abs/2018ApJ...863..171S} {863, 171}

\bibitem[\protect\citeauthoryear{{Sills}, {Rieder}, {Scora}, {McCloskey}  \&
  {Jaffa}}{{Sills} et~al.}{2018}]{sills18}
{Sills} A.,  {Rieder} S.,  {Scora} J.,  {McCloskey} J.,   {Jaffa} S.,  2018,
  \mnras, \href {https://ui.adsabs.harvard.edu/abs/2018MNRAS.477.1903S} {477,
  1903}

\bibitem[\protect\citeauthoryear{{Smilgys} \& {Bonnell}}{{Smilgys} \&
  {Bonnell}}{2017}]{sb17}
{Smilgys} R.,  {Bonnell} I.~A.,  2017, \mn@doi [\mnras]
  {10.1093/mnras/stx2396}, \href
  {https://ui.adsabs.harvard.edu/abs/2017MNRAS.472.4982S} {472, 4982}

\bibitem[\protect\citeauthoryear{{Smith}, {Fellhauer}, {Goodwin}  \&
  {Assmann}}{{Smith} et~al.}{2011}]{smith11}
{Smith} R.,  {Fellhauer} M.,  {Goodwin} S.,   {Assmann} P.,  2011, \mn@doi
  [\mnras] {10.1111/j.1365-2966.2011.18604.x}, \href
  {https://ui.adsabs.harvard.edu/abs/2011MNRAS.414.3036S} {414, 3036}

\bibitem[\protect\citeauthoryear{{Toonen}, {Nelemans}  \& {Portegies
  Zwart}}{{Toonen} et~al.}{2012}]{seba2}
{Toonen} S.,  {Nelemans} G.,   {Portegies Zwart} S.,  2012, \mn@doi [\aap]
  {10.1051/0004-6361/201218966}, \href
  {https://ui.adsabs.harvard.edu/abs/2012A&A...546A..70T} {546, A70}

\bibitem[\protect\citeauthoryear{{Tress}, {Sormani}, {Glover}, {Klessen},
  {Battersby}, {Clark}, {Hatchfield}  \& {Smith}}{{Tress}
  et~al.}{2020}]{tress20}
{Tress} R.~G.,  {Sormani} M.~C.,  {Glover} S. C.~O.,  {Klessen} R.~S.,
  {Battersby} C.~D.,  {Clark} P.~C.,  {Hatchfield} H.~P.,   {Smith} R.~J.,
  2020, \mn@doi [\mnras] {10.1093/mnras/staa3120}, \href
  {https://ui.adsabs.harvard.edu/abs/2020MNRAS.499.4455T} {499, 4455}

\bibitem[\protect\citeauthoryear{{Tre{\ss}}, {Sormani}, {Smith}, {Glover},
  {Klessen}, {Mac Low}, {Clark}  \& {Duarte-Cabral}}{{Tre{\ss}}
  et~al.}{2021}]{tress21}
{Tre{\ss}} R.~G.,  {Sormani} M.~C.,  {Smith} R.~J.,  {Glover} S. C.~O.,
  {Klessen} R.~S.,  {Mac Low} M.-M.,  {Clark} P.,   {Duarte-Cabral} A.,  2021,
  \mn@doi [\mnras] {10.1093/mnras/stab1683}, \href
  {https://ui.adsabs.harvard.edu/abs/2021MNRAS.505.5438T} {505, 5438}

\bibitem[\protect\citeauthoryear{{Tsuge}, {Tachihara}, {Fukui}, {Sano},
  {Tokuda}, {Ueda}  \& {Iono}}{{Tsuge} et~al.}{2021}]{tsuge21}
{Tsuge} K.,  {Tachihara} K.,  {Fukui} Y.,  {Sano} H.,  {Tokuda} K.,  {Ueda} J.,
    {Iono} D.,  2021, \mn@doi [\pasj] {10.1093/pasj/psab008}, \href
  {https://ui.adsabs.harvard.edu/abs/2021PASJ...73..417T} {73, 417}

\bibitem[\protect\citeauthoryear{{Urquhart} et~al.,}{{Urquhart}
  et~al.}{2021}]{urquhart21}
{Urquhart} J.~S.,  et~al., 2021, \mn@doi [\mnras] {10.1093/mnras/staa2512},
  \href {https://ui.adsabs.harvard.edu/abs/2021MNRAS.500.3050U} {500, 3050}

\bibitem[\protect\citeauthoryear{{Van der Helm}, {Saladino}, {Portegies Zwart}
  \& {Pols}}{{Van der Helm} et~al.}{2019}]{stellarwindpy}
{Van der Helm} E.,  {Saladino} M.~I.,  {Portegies Zwart} S.,   {Pols} O.,
  2019, \mn@doi [\aap] {10.1051/0004-6361/201732020}, \href
  {https://ui.adsabs.harvard.edu/abs/2019A&A...625A..85V} {625, A85}

\bibitem[\protect\citeauthoryear{{V{\'a}zquez-Semadeni},
  {Gonz{\'a}lez-Samaniego}  \& {Col{\'\i}n}}{{V{\'a}zquez-Semadeni}
  et~al.}{2017}]{vazquezsemadeni17}
{V{\'a}zquez-Semadeni} E.,  {Gonz{\'a}lez-Samaniego} A.,   {Col{\'\i}n} P.,
  2017, \mn@doi [\mnras] {10.1093/mnras/stw3229}, \href
  {https://ui.adsabs.harvard.edu/abs/2017MNRAS.467.1313V} {467, 1313}

\bibitem[\protect\citeauthoryear{{Vera}, {Alonso}  \& {Coldwell}}{{Vera}
  et~al.}{2016}]{vera16}
{Vera} M.,  {Alonso} S.,   {Coldwell} G.,  2016, \mn@doi [\aap]
  {10.1051/0004-6361/201628750}, \href
  {https://ui.adsabs.harvard.edu/abs/2016A&A...595A..63V} {595, A63}

\bibitem[\protect\citeauthoryear{{Vogel}, {Kulkarni}  \& {Scoville}}{{Vogel}
  et~al.}{1988}]{vogel88}
{Vogel} S.~N.,  {Kulkarni} S.~R.,   {Scoville} N.~Z.,  1988, \mn@doi [\nat]
  {10.1038/334402a0}, \href
  {https://ui.adsabs.harvard.edu/abs/1988Natur.334..402V} {334, 402}

\bibitem[\protect\citeauthoryear{{Vutisalchavakul}, {Evans}  \&
  {Heyer}}{{Vutisalchavakul} et~al.}{2016}]{vutis16}
{Vutisalchavakul} N.,  {Evans} Neal~J. I.,   {Heyer} M.,  2016, \mn@doi [\apj]
  {10.3847/0004-637X/831/1/73}, \href
  {https://ui.adsabs.harvard.edu/abs/2016ApJ...831...73V} {831, 73}

\bibitem[\protect\citeauthoryear{{Wall}, {McMillan}, {Mac Low}, {Klessen}  \&
  {Portegies Zwart}}{{Wall} et~al.}{2019}]{wall19}
{Wall} J.~E.,  {McMillan} S. L.~W.,  {Mac Low} M.-M.,  {Klessen} R.~S.,
  {Portegies Zwart} S.,  2019, \mn@doi [\apj] {10.3847/1538-4357/ab4db1}, \href
  {https://ui.adsabs.harvard.edu/abs/2019ApJ...887...62W} {887, 62}

\bibitem[\protect\citeauthoryear{{Wang}, {Iwasawa}, {Nitadori}  \&
  {Makino}}{{Wang} et~al.}{2020}]{petarcode}
{Wang} L.,  {Iwasawa} M.,  {Nitadori} K.,   {Makino} J.,  2020, \mn@doi
  [\mnras] {10.1093/mnras/staa1915}, \href
  {https://ui.adsabs.harvard.edu/abs/2020MNRAS.497..536W} {497, 536}

\bibitem[\protect\citeauthoryear{{Wright} \& {Mamajek}}{{Wright} \&
  {Mamajek}}{2018}]{wm18}
{Wright} N.~J.,  {Mamajek} E.~E.,  2018, \mn@doi [\mnras]
  {10.1093/mnras/sty207}, \href
  {https://ui.adsabs.harvard.edu/abs/2018MNRAS.476..381W} {476, 381}

\bibitem[\protect\citeauthoryear{{Yu}, {de Grijs}  \& {Chen}}{{Yu}
  et~al.}{2011}]{yu11}
{Yu} J.,  {de Grijs} R.,   {Chen} L.,  2011, \mn@doi [\apj]
  {10.1088/0004-637X/732/1/16}, \href
  {https://ui.adsabs.harvard.edu/abs/2011ApJ...732...16Y} {732, 16}

\bibitem[\protect\citeauthoryear{{Yu}, {Ho}  \& {Wang}}{{Yu}
  et~al.}{2021}]{yu21}
{Yu} S.-Y.,  {Ho} L.~C.,   {Wang} J.,  2021, \mn@doi [\apj]
  {10.3847/1538-4357/ac0c77}, \href
  {https://ui.adsabs.harvard.edu/abs/2021ApJ...917...88Y} {917, 88}

\bibitem[\protect\citeauthoryear{{de Zeeuw}, {Hoogerwerf}, {de Bruijne},
  {Brown}  \& {Blaauw}}{{de Zeeuw} et~al.}{1999}]{dezeeuw99}
{de Zeeuw} P.~T.,  {Hoogerwerf} R.,  {de Bruijne} J.~H.~J.,  {Brown} A.~G.~A.,
   {Blaauw} A.,  1999, \mn@doi [\aj] {10.1086/300682}, \href
  {https://ui.adsabs.harvard.edu/abs/1999AJ....117..354D} {117, 354}

\makeatother
\end{thebibliography}


%
%

%

\bsp	
\label{lastpage}
\end{document}